\documentclass{sigma}
\usepackage{amssymb}
\usepackage{bbm}

\newcommand{\Z}{\mathbbm Z}

\newcommand{\eqb}{\begin{equation}}
\newcommand{\eqe}{\end{equation}}
\newcommand{\dmb}{\begin{displaymath}}
\newcommand{\dme}{\end{displaymath}}
\newcommand{\pd}{\partial}

\newcommand{\eab}{\begin{eqnarray}}
\newcommand{\eae}{\end{eqnarray}}

\newcommand{\e}{\mbox{e}}
\newcommand{\be}{\begin{equation}}
\newcommand{\ee}{\end{equation}}

\begin{document}

\renewcommand{\PaperNumber}{001}

\FirstPageHeading

\ShortArticleName{Yang-Mills thermodynamics}
\ArticleName{Yang-Mills thermodynamics}
\Author{Ralf Hofmann}
\AuthorNameForHeading{R. Hofmann}
\Address{Institut f\"ur Theoretische Physik, 
Universit\"at Karlsruhe (TH),  
Kaiserstr. 12, 76131 Karlsruhe, Germany}
\EmailD{hofmann@particle.uni-karlsruhe.de}
\URLaddressD{http://www.thphys.uni-heidelberg.de/$\sim$hofmann/ }
\ArticleDates{Received October 4, 2007, in final form ????; Published online ????}

\Abstract{We present a quantitative analysis 
of Yang-Mills thermodynamics in 4D flat spacetime. The focus is on the gauge group 
SU(2). Results for SU(3) are mentioned 
in passing. Although all essential arguments and results were reported
elsewhere we summarize them here in a concise way and offer 
a number of refinements and some additions.}

\Keywords{holonomy; calorons; Polyakov loop; adjoint Higgs mechanism; spatial coarse-graining;
deconfinement; renormalizability; Legendre transformation;
unitary-Coulomb gauge; maximal resolution; 
loop expansion; irreducible bubble diagram; BPS monopole; monopole condensate; 
abelian Higgs mechanism; dual gauge field; preconfinement; center-vortex loop; 
spin-1/2 fermion; 't Hooft loop; Hagedorn transition; condensate of paired 
center-vortex loops; total confinement; asymptotic series; $\lambda\phi^4$-theory in 1D; 
Borel summability and analytic continuation}

\Classification{70S15; 74A15; 82B10; 82B28} 

\section{Introduction}

It was Planck who first demonstrated the power of statistical 
methods in quantitatively understanding a gauge theory \cite{Planck}. 
His important suggestion was to subject indeterministic phase and amplitude changes 
of a single resonator in the wall of a cavity -- 
in thermal equilibrium with the contained electromagnetic radiation --  
to an averaging procedure dictated by the laws of (statistical) 
thermodynamics. Appealing to a known, classically derived 
result on gross features of the so-called black-body spectrum (Wien's displacement law), 
postulating a partitioning of the total energy into multiples of a smallest 
unit, and appealing to Boltzmann's statistical definition of entropy, 
Planck deduced his famous radiation law. As an aside, he discovered a 
universal quantum of action needed to relate the 
entropy (disorder) and mean energy of a single 
resonator to its frequency. The robustness of his 
result is demonstrated by the fact that even for physical objects
sizably deviating from ideal black bodies Planck's radiation law holds 
to a high degree of accuracy. 

The purpose of the present article is to give a concise presentation 
of results, accumulated over the last four years,  
on generalizations of the thermal U(1) gauge theory studied by Planck: SU(2)
and SU(3) Yang-Mills thermodynamics. It is possible that 
an SU(2) gauge symmetry, dynamically broken down to U(1) by a 
deconfining thermal ground state, {\sl underlies} photon propagation 
\cite{Hofmann2005-1,GH2005,SHG2006-1,SHG2006-2,SHGS2007}. 
We do not here consider SU(N) gauge theories with $\infty\ge$\,N\,$\ge 4$ the 
reason\footnote{The case N$=\infty$ may be an important exception
  \cite{Hofmann2005L}. The nonuniqueness of the phase 
diagram is due to an incomplete Abelianization of the fundamental gauge
symmetry by an adjoint Higgs field: There is no 
principle which decides at what 
temperature a nonabelian factor is broken to its Abelian subgroup.} 
being nonunique phase diagrams \cite{Hofmann2005L,Hofmann2005S}. 

Yang-Mills thermodynamics strongly relates to the concept of 
emergent phenomena. On the most basic level, temperature itself is an
emergent phenomenon depending on the fluctuating degrees of freedom
defining it. Conversely, the mass of a magnetic monopole, which, as a 
short-lived field configuration contributes to the thermodynamics of the
Yang-Mills ground state at high temperatures, is determined by 
temperature. That is, a single monopole owes its existence to the 
existence of all other fluctuating monopoles and
antimonopoles in the ensemble. 

In Yang-Mills thermodynamics various temperature-dependent 
emergent phenomena, facilitated by 
topologically nontrivial mappings from submanifolds of four dimensional 
spacetime into the (sub)manifold(s) of the gauge group,
dominate the ground-state thermodynamics in three different phases. Albeit their microscopic
dynamics is complex and not accessible to analytic treatment, a
thermodynamically implied spatial coarse-graining down 
to a uniquely determined resolution, determined by the Yang-Mills scale and temperature, yields 
accurate and technically managable representations of 
the (ultraviolet-regulated) partition function at any given temperature\footnote{An exception is the 
nonthermal behavior shortly below the Hagedorn transition.}. Here the
term spatial coarse-graining refers to the 
process of eliminating short-wavelength gauge-field 
fluctuations in favor for effective fields and their 
couplings in a reformulation of the same 
partition function at lower and lower resolution and at a 
given temperature. This coarse-graining leads to the emergence of an effective 
action $S_{\tiny\mbox{eff}}$ being a functional of
the effective fields which determines the weight for the functional integration over 
the latter in the reformulated partition function valid at a given
maximal resolution. 

Remarkably, starting out from exact BPS saturated 
solutions to the euclidean field equations in the deconfining phase, the derivation of the thermal 
ground state in the deconfining, high-temperature phase 
makes no reference to the way of how the continuum partition 
function of the theory is regularized in the ultraviolet. 
This is a consequence of the possibility to compute the 
expectation value of a uniquely determined operator, representing the
phase $\hat{\phi}$ of an adjoint scalar field $\phi$, over suitable, topologically nontrivial, and BPS 
saturated configurations {\sl first} and to {\sl subsequently} investigate 
its {\sl average} effect on {\sl integrated} fluctuations and its 
{\sl direct} effect on {\sl explicit} field configurations 
in the topologically trivial sector. No 
reference is made to an ultraviolet regularization in the process. When
decreasing the maximal resolution in the BPS saturated situation 
it is possible to deduce definite information about $\phi$'s modulus 
$|\phi|$ from its phase $\hat{\phi}$ since the process of spatial
coarse-graining, which for $\hat{\phi}$ rapidly saturates the
{\sl limit of vanishing} maximal resolution at a {\sl finite} maximal
resolution, yields a spacetime homogeneous (and of course isotropic) value\footnote{Due to the nonfluctuating nature of the 
field $\phi$ or higher Lorentz-spin fields, potentially generated by a coarse-graining 
over the BPS sector of the fundamental theory, infinite-volume thermodynamics excludes the existence of the latter and 
demands homogeneity (constancy) of the former's gauge invariant modulus.} of $|\phi|$ starting
from a maximal resolution determined by $|\phi|$. 

That is, spatial coarse-graining over noninteracting, BPS saturated
configurations in the (only admissible) sector with topological charge
modulus $|Q|=1$ dimensionally formally  reduces this sector of the theory 
from $D=4$ to $D=1$ (quantum mechanics with periodic-in-time
trajectories) at a finite maximal resolution. Thermodynamically, the gauge-invariant quantity 
$|\phi|$ must not carry energy (or momentum in 4D) because of its inherited BPS saturation. 
This is also the reason why no local vertices of the field $\phi$ with 
the topologically trivial, coarse-grained gauge field involving three of more external legs of the latter may exist: 
Such vertices would on the level of the effective theory convey energy-momentum transfer 
from the topologically trivial to the topologically nontrivial but BPS saturated sector 
of the theory which contradicts the very existence of the field $\phi$. 
Notice, however, that on the {\sl fundamental} level topologically trivial 
fluctuations do interact with the topologically nontrivial sector 
exchanging energy-momentum associated with a resolution higher 
than $|\phi|$ thus having no visible effect on the field $\phi$. These interactions, introducing a 
temporary (anti)caloron holonomy in addition to short-range radiative corrections, 
are described by a pure-gauge configuration in the effective theory which in fact lifts the energy density of the 
thermal ground-state from zero to a finite positive value.  

Despite 
the fact that the Stefan-Boltzmann limit is approached in a power-like
and thus rapid way with increasing temperature the 
inherent resolution $|\phi|$ of the thermal Yang-Mills system decreases 
with temperature, and thus any memory of the
short-distance regularization of the partition function is wiped out, see also 
\cite{PerezArroyoMontero} for the corresponding lattice 
observation (increasing delocalization of 
topological charge with increasing temperature). 
Conceptually, this is in accord with the situation known in
zero-temperature perturbation theory 
where our ignorance about the values and UV regularization dependence of bare parameters is 
shown to be no obstacle to the predictivity of the quantum Yang-Mills 
theory since, upon their dressing at a finite resolution, only finitely many such parameters need to 
be fixed \cite{'tHooftVeltman-1,'tHooftVeltman-2,'tHooftVeltman-3,'tHooftVeltman-4}. 

Following the appreciated advice of a Referee, 
the presentation in this paper resorts to a hybrid style: A statement,
whose validity is argued for in a more physical rather than rigorous 
mathematical way, and a number of definitions are highlighted by 
slanted script and are interspersed into the argumentation, a statement whose
validity under stated assumptions is verified by
{\sl direct calculation} is presented according to custom within the 
mathematics literature, apologies to the irritated reader. But even though the subject 
presented (4D quantum field theory) still awaits its rigorous 
mathematical foundation the pragmatic line of pursuit followed in the
present work unearthens a number of unexpected, quantitatively very
precisely representable facts which, as the author is convinced of, do not
depend on a future, rigorous formulation of quantum field theory.    

For maximal benefit it is recommended to read the present
article in conjunction with \cite{Hofmann2005L}. That article contains 
extended discussions of the involved
physics, explicit expressions for the 
relevant topological field configurations, and graphic 
displays of numerical results. Although the use of differential forms
would simplify certain statements in the beginning of Sec.\,\ref{DP} 
our presentation entirely resorts to the component
notation. 

The following conventions will be used: Einstein summation (summation 
over doubly occurring indices), solely 
lower-case indices for contractions in the euclidean formulation, 
lower-case and upper-case
indices for contractions in the real-time formulation, and natural 
units ($\hbar=k_B=c=1$).      

The outline of this work is as follows: In Sec.\,\ref{EFS} we remind the 
reader of basic facts about thermal Yang-Mills gauge field theory. Sec.\,\ref{DP}
discusses the deconfining phase where a thermal ground 
state emerges upon a spatial coarse-graining over interacting, BPS
saturated field configurations and where the Yang-Mills scale
occurs as a purely nonperturbative constant of integration. We give
tight estimates on the goodness of the finite-volume 
saturation of the coarse-graining process, and we account for the 
radiative corrections in the effective theory. 
The thermodynamics of the intermediate, preconfining phase is addressed
in Sec.\,\ref{PP}. Here the unbroken abelian gauge 
symmetry of the deconfining phase is dynamically broken 
by monopole-antimonopole condensate(s). Emphasis is put on a discussion of 
supercooling which takes place because the switch from small to large 
caloron/anticaloron holonomy is energetically disfavored. 
Finally, in Sec.\,\ref{CP} we elucidate the process of the decay of the
monopole-antimonopole condensate of the preconfining 
phase giving rise to a zero-pressure and zero-energy density ground state
in the confining phase. In that phase no propagating gauge modes 
exist, and the spectrum is represented by single or selfintersecting
center-vortex loops which we interprete as spin-1/2 fermions. 
The naive series for the thermodynamic pressure represents an 
asymptotic expansion, and we show its Borel summability for complex values
of the expansion parameter. Upon continuation to the 
physical regime a sign-indefinite imaginary part is 
encountered which for sufficiently small temperatures, however, is, 
smaller than the definite real part.

\section{Thermal Yang-Mills theory\label{EFS}}

\subsection{Euclidean formulation and symmetries}

{\sl On a flat, four-dimensional euclidean spacetime with coordinates
$0\le\tau\le\beta\equiv T^{-1}$ (time) and $\vec{x}$ (infinite
three-dimensional space) 
the partition function $Z$ of a pure Yang-Mills 
gauge-field theory subject to the gauge group SU(N) is formally defined as 
\begin{equation}
\label{fundpartfunct}
Z\equiv \int_{A_\mu(\tau=0,\vec{x})=A_\mu(\tau=\beta,\vec{x})} {\cal D} A_\mu\,\exp[-S]\,,  
\end{equation} 
where the gauge-field configuration $A_\mu$ is Lie-algebra valued, $A_\mu\equiv A_\mu^a\,t_a\,,\ \
(a=1,\cdots,\mbox{N}^2-1)$, with the generators $t_a$ in the fundamental
representation normalized as
$\mbox{tr}\,t_at_b=\frac12\delta_{ab}\,,$ and $T$ is the temperature. 
The action $S$ is defined as 
$S\equiv\frac{1}{2g^2}\,\mbox{tr}\,\int_0^\beta d\tau\int
d^3x\,F_{\mu\nu}F_{\mu\nu}$ where $g$ is a dimensionless coupling constant, $F_{\mu\nu}=\pd_\mu
A_\nu-\pd_\nu A_\mu-i[A_\mu,A_\nu]$\,,\ ($\mu,\nu=1,\cdots,4$), and the measure of (functional) integration is ${\cal D}
A_\mu\equiv\prod_{\tau,\vec{x},a} dA^a_\mu(\tau,\vec{x})$. Here the product is
over the continuously varying values of the coordinates 
$\tau$, $\vec{x}$, and over $a=1,\cdots,\mbox{N}^2-1$}\vspace{0.2cm}\\ 
For the gauge group SU(2) we set $t_a=\frac12\lambda_a$ where
$\lambda_a$ are the Pauli matrices. The integration measure ${\cal D}
A_\mu\equiv\prod_{\tau,\vec{x}} dA_\mu(\tau,\vec{x})$ is ill-defined as
it stands. If the field theory is endowed with an ultraviolet regularization 
then the infinite product is over a discrete index. 
We will argue that Yang-Mills thermodynamics, formally defined
\footnote{This partition function implies Legendre transformations between 
formally defined thermodynamical quantities like pressure (minus free energy), 
energy density, and entropy density.} by Eq.\,(\ref{fundpartfunct}),
does not relate to the way how sense is made of the formal object in
Eq.\,(\ref{fundpartfunct}) by introducing a minimal length as
long as the selfadjusting resolution in the system is lower than the
resolution associated with this ultraviolet scale. 
As already discussed in the
Introduction at a given temperature $T$ a unique 
maximal resolution $|\phi|$ appears to emerge. Since $|\phi|$ 
decays like a power when increasing $T$ it is guaranteed that 
the high-temperature physics is insensitive to 
any definite ultraviolet substantialization of
Eq.\,(\ref{fundpartfunct}). 
The action density $\frac{1}{2g^2}\,\mbox{tr}\,F_{\mu\nu}F_{\mu\nu}$ in
Eq.\,(\ref{fundpartfunct}) is invariant under gauge transformations
$A_\mu\stackrel{\Omega}\longrightarrow\Omega A_\mu\Omega^\dagger+i\Omega\pd_\mu\Omega^\dagger$, where
$\Omega$ is an element of SU(N) in the fundamental representation, but 
the functional integration is carried out over gauge-inequivalent, 
periodic-in-$\tau$ gauge-field configurations. 

In lattice definitions of the partition function in 
Eq.\,(\ref{fundpartfunct}) one can show its invariance under temporally 
local center transformations, that is, under gauge transformations which are 
periodic up to a multiplication with a center element: $\Omega(\tau=0,\vec{x})=Z\,\Omega(\tau=\beta,\vec{x})$ where 
$Z\in {\mathbbm Z}_{\tiny\mbox{N}}$ and thus the relevant gauge group
actually is SU(N)$/{\mathbbm Z}_{\tiny\mbox{N}}$. If this (electric) 
center symmetry of the Yang-Mills action $S$, subjected to the gauge group
SU(N), is broken dynamically then a ${\mathbbm Z}_{\tiny\mbox{N}}$
degeneracy of the ground state must be present. 
Furthermore, the action $S$ is invariant under continuous spatial rotations and
translations. There is also an invariance w.r.t. time translations
$\tau\to\tau+\tau_0$, $0\le\tau_0\le\beta$.

\subsection{BPS saturated field configurations at finite temperature}

The Euler-Lagrange equations, $D_\mu F_{\mu\nu}=0$ (stationarity of the action,  
$\frac{\delta S}{\delta A_\mu}=0$) are solved by configurations $A_\mu$ 
obeying the (anti)selfduality condition $F_{\mu\nu}=\pm\tilde{F}_{\mu\nu}$. Here
$D_\mu\cdot=\pd_\mu\cdot-i[A_\mu,\cdot]$ and
$\tilde{F}_{\mu\nu}\equiv\frac12\epsilon_{\mu\nu\kappa\lambda}\,F_{\kappa\lambda}$
where $\epsilon_{\mu\nu\kappa\lambda}$ is the totally antisymmetric
tensor with $\epsilon_{1234}=1$. (Anti)selfdual configurations saturate
the Bogomol'nyi bound on the action (BPS saturation):
$S=\frac{8\pi^2}{g^2}|Q|$ where $Q\equiv\frac{1}{32\pi^2}\int_0^\beta
d\tau\int d^3x\,F^a_{\mu\nu}\tilde{F}^a_{\mu\nu}\in\Z$ is the 
topological charge. $Q$ is finite and quantized if according boundary 
conditions are imposed. At finite temperature we 
consider BPS saturated, finite-action configurations $A_\mu$ which
behave accordingly at spatial infinity and are periodic in 
$\tau$. In this case $Q$ is an integer. As is usual, we 
refer to these configurations as calorons/anticalorons for a positive/negative sign of
$Q$. (The presence of spatial boundaries, which are not at infinity, 
would explicitely break the translational invariance. We so 
far have no well backed up insight on how to treat Yang-Mills thermodynamics 
analytically in this case which, in a softend fashion, is also conveyed by 
external sources. A possibility for the treatment of mild distortions 
is to deform the undistorted situation adiabatically by letting 
$T\to T(\vec{x})$ and/or $\Lambda\to \Lambda(\vec{x})$ where $\Lambda$
denotes the Yang-Mills scale). Solutions to the 
Euler-Lagrange equations, which are not (anti)selfdual, have been
constructed numerically, for recent work on axially symmetric
finite-temperature and instanton/antiinstanton configurations and see
\cite{Shnir} and \cite{Radu}, respectively.         
\begin{proposition}
\label{emtens}
The euclidean energy-momentum tensor $\theta_{\mu\nu}\equiv -F^a_{\mu\lambda}F^a_{\nu\lambda}+
\frac14\delta_{\mu\nu}\,F^a_{\kappa\lambda}F^a_{\kappa\lambda}$ and
every local, scalar composite of the form
tr\,$t^a\,F_{\mu\kappa}F_{\kappa\mu}$,
tr\,$t^a\,F_{\mu\kappa}F_{\kappa\nu}F_{\nu\mu}$,
tr\,$t^a\,F_{\mu\kappa}F_{\kappa\nu}F_{\nu\lambda}F_{\lambda\mu},
\cdots$ vanish identically on a BPS saturated field configuration $A_\mu$.
\end{proposition}
\begin{proof}
Routine computation.  
\end{proof}
Notice that for $Q=0$ we have $S=0$ implying
that calorons in this sector are pure gauges: $A_\mu=i\Omega\pd_\mu\Omega^\dagger$.\vspace{0.2cm}\\ 
{\sl The Polyakov loop $P(\vec{x})[A]$ is defined as $P(\vec{x})[A]\equiv{\cal P}\exp[i\int_0^\beta
A_4(\tau,\vec{x})]$ where the symbol ${\cal P}$ demands
path-ordering.}\vspace{0.2cm}\\ 
{\sl A caloron/anticaloron is said to be of trivial 
holonomy if\\  
$P_\infty[A]\equiv\lim_{|\vec{x}|\to\infty}P(\vec{x})[A]\in\mbox{center of the
  gauge group}$.\\ \noindent (Recall that for SU(N): center=$\{\exp[2\pi
i\frac{k}{\tiny\mbox{N}}]{\mathbbm 1}_{\tiny\mbox{N}}|
k=0,1,\cdots,\mbox{N}-1\}={\mathbbm
  Z}_{\tiny\mbox{N}}$.)}\vspace{0.2cm}\\ 
Notice that for the gauge group SU(N) this definition does not depend on the choice of
gauge as long as $\Omega(\tau=0,\vec{x})=Z\Omega(\tau=\beta,\vec{x})$, where 
$Z\in {\mathbbm Z}_{\tiny\mbox{N}}$, since 
$P_\infty[A]\stackrel{\Omega}\longrightarrow\Omega(\tau=0) P_\infty[A]\Omega^\dagger(\tau=\beta)\,.$ 
\begin{example}
\label{HarringtonShepard} 
For the gauge group SU(2) the following calorons/anticalorons (Harrington-Shepard (HS) 
\cite{HS1977}) are of trivial holonomy and of
topological charge $Q=\pm 1$: 
\begin{equation}
A^C_\mu(\tau,\vec{x})=\bar{\eta}^a_{\mu\nu} t_a \pd_{\nu}\ln
\Pi(\tau,r)\ \ \ \ (\mbox{caloron}, Q=+1)\,,\nonumber\\ 
\end{equation}
\begin{equation}
A^A_\mu(\tau,\vec{x})=\eta^a_{\mu\nu} t_a
\pd_{\nu}\ln \Pi(\tau,r)\ \ \ \ (\mbox{anticaloron}, Q=-1)\,, 
\end{equation}
where the 't Hooft symbols $\eta_{a\mu\nu}$ and $\bar{\eta}^a_{\mu\nu}$ 
are defined as $\eta^a_{\mu\nu}=\epsilon^a_{\mu\nu} +
\delta^a_{\mu}\delta_{\nu4} - \delta^a_{\nu}\delta_{\mu4}$ and 
$\bar\eta^a_{\mu\nu}=\epsilon^a_{\mu\nu} - \delta^a_{\mu}\delta_{\nu4}
+ \delta^a_{\nu}\delta_{\mu4}$, and the prepotential $\Pi$ is given as 
\begin{equation*}
\Pi(\tau,r)\equiv1+\frac{\pi\rho^2}{\beta r}
\frac{\sinh\left(\frac{2\pi r}{\beta}\right)}{\cosh\left(\frac{2\pi r}{\beta}\right)-
\cos\left(\frac{2\pi\tau}{\beta}\right)}\,,\ (r\equiv|\vec{x}|)\,.
\end{equation*}
The dimensionful modulus $\rho$ is inherited from the 
singular-gauge instanton configuration with 
prepotential $\Pi_0(\tau,r)=1+\frac{\rho^2}{\tau^2+\vec{x}^2}$ since
$\Pi$ is obtained from $\Pi_0$ by superimposing its infinitely many images
in mirrors placed at $\tau=0$ and $\tau=\beta$ to generate periodicity
in $\tau$. Additional moduli are the shifts $\tau\to\tau+\tau_z\,,\
(0\le\tau_z\le\beta)\,,$ and $\vec{x}\to\vec{x}+\vec{z}$ and, if one
wishes, global gauge transformations.  
\end{example}
\begin{example}
\label{LLKvB}
For the gauge group SU(2) there exist
\cite{NahmP,NahmL,NahmD-1,NahmD-2,Atiyah} explicitely constructed 
\cite{LeeLu1998,KraanVanBaalNPB1998-1,
KraanVanBaalNPB1998-2,KraanVanBaalNPB1998-3,Brower1998} BPS
calorons/anticalorons (Lee-Lu-Kraan-van-Baal (LLKvB)) of nontrivial holonomy and
topological charge $Q=\pm 1$. For the trivial-holonomy case
configurations with $|Q|>1$ were constructed in
\cite{Actor1983,Chkrabarti1987}. 

In contrast to their trivial-holonomy
counterparts \cite{GrossPisarskiYaffe} these configurations are 
in isolation instable under quantum deformation \cite{Diakonov2004}: For a holonomy
sufficiently close to trivial and for $\rho>0$ the static BPS magnetic monopole and
antimonopole constituents \cite{tHooft1974,Polayakov1974,PrasadSommerfield1974}
attract under the influence of quantum fluctuations and thus eventually annihilate
one another. This relaxes the LLKvB caloron or anticaloron back to the
stable situation of a HS caloron or anticaloron. For a holonomy far from trivial and for $\rho>0$
BPS magnetic monopole and antimonopole repulse \cite{Diakonov2004} one another under the
influence of quantum fluctuations. As a consequence, the large-holonomy LLKvB caloron or 
anticaloron dissociates into a pair of a screened BPS magnetic monopole and its
antimonopole. Screening occurs due to the presence of short-lived
magnetic dipoles that are provided by {\sl intermediary} small-holonomy LLKvB
calorons and anticalorons. 

In the seminal work \cite{GrossPisarskiYaffe}
nontrivial-holonomy calorons/anticalorons were argued to not contribute
to the partition function based on the observation that 
quantum corrections produce a term in their effective action which 
diverges like the three-volume of the system. This argument 
is certainly correct if holonomy is considered a quantity that
is {\sl externally sustained} at a definite value. Hence no {\sl explicit} integration
over the holonomy must occur in any first-principle 
evaluation of the (ultraviolet regularized) partition function at
sufficiently large temperature. However, viewed as a dynamical,
short-lived quantity nontrivial holonomy {\sl does} occur through the
quantum induced deformation of the trivial-holonomy case. Partially based on
theoretical work on fermionic zero modes, used as a diagnostics for lumps of 
topological charge \cite{BruckmannII,BruckmannIII} to avoid the
application of a cooling procedure to a given configuration, this is impressively
demonstrated by many lattice investigations. A nonexhaustive list of
references is \cite{PerezArroyoMontero,IlgenfritzI,IlgenfritzII,IlgenfritzIII,
GattringerI,GattringerII}. 

The construction of nontrivial-holonomy calorons/anticalorons of higher 
topological charge for a Yang-Mills theory subject to the gauge group
SU(N) was investigated by Bruckmann and van Baal in \cite{BruckmannI} and the 
explicit form of the solution was given for $|Q|=2$. The interesting 
result is that these calorons possess $n|Q|$ constituents 
monopoles whose sum of magnetic charges (with respect to
U(1)$^{\tiny\mbox{N-1}}$) is nil. Recently, a nontrivial-holonomy caloron
of $Q=2$ and nonvanishing, overall magnetic charge was constructed \cite{Harland}.                   
\end{example}
\begin{remark}
\label{temperatureLLKvB}
On the level of the euclidean saddlepoint one has 
for the masses $m_1$ of a BPS magnetic monopole and $m_2$ of its
antimonopole inside a LLKvB caloron: 
$m_1+m_2=8\pi^2 T$ \cite{LeeLu1998}. Thus already on the classical level one 
observes the remarkable fact that 
the emergence of a particular 
monopole depends on the emergence of temperature or in other words on
the existence of all other fluctuating monopoles and antimonopoles 
and propagating gauge fields in the ensemble.      
\end{remark}

\subsection{Propagating fields at finite temperature: Q=0\label{toptriv}}

In a given gauge and in the euclidean formulation 
the topologically trivial sector $\{\delta A_\mu\}$ is represented by a
superposition of plane waves:
\eqb
\label{fourierb}
\delta A_\mu(\tau,{\vec x})=\sum_{n=-\infty}^{n=\infty} 
\exp\left[2\pi in \frac{\tau}{\beta}\right]\,\delta\bar{A}_{\mu,n}({\vec x})\,
\eqe
where $\delta\bar{A}_{\mu,n}({\vec x})=\int
d^3k\,\alpha_{\mu,n}(\vec{k})\,\exp[i\vec{k}\cdot\vec{x}]$ and the
function $\alpha_{\mu,n}(\vec{k})$ falls off sufficiently fast in
$|\vec{k}|$ and in $n$ for the integrals and the sum in
Eq.\,(\ref{fourierb}) to exist, respectively. The quantity
$\frac{2\pi n}{\beta}$ is called $n$th Matsubara frequency. 
\begin{remark} 
Upon a Wick rotation $\tau\to it\,,\ \ (t\ \mbox{real})\,,$ one
shows that the propagator of the field $\delta A_\mu$ decomposes 
into a quantum part (describing a particle of four-momentum $p$ possibly 
being off the mass shell, $p^2\equiv p^\mu p_\mu\not=0$) 
and a thermal part (describing thermalized on-shell propagation), 
see for example \cite{Kapusta}.  
\end{remark} 

\section{Deconfining phase\label{DP}}

\subsection{Thermal ground state: Interacting calorons/anticalorons}

\begin{figure}[tb]
\centering
\includegraphics[width=14cm]{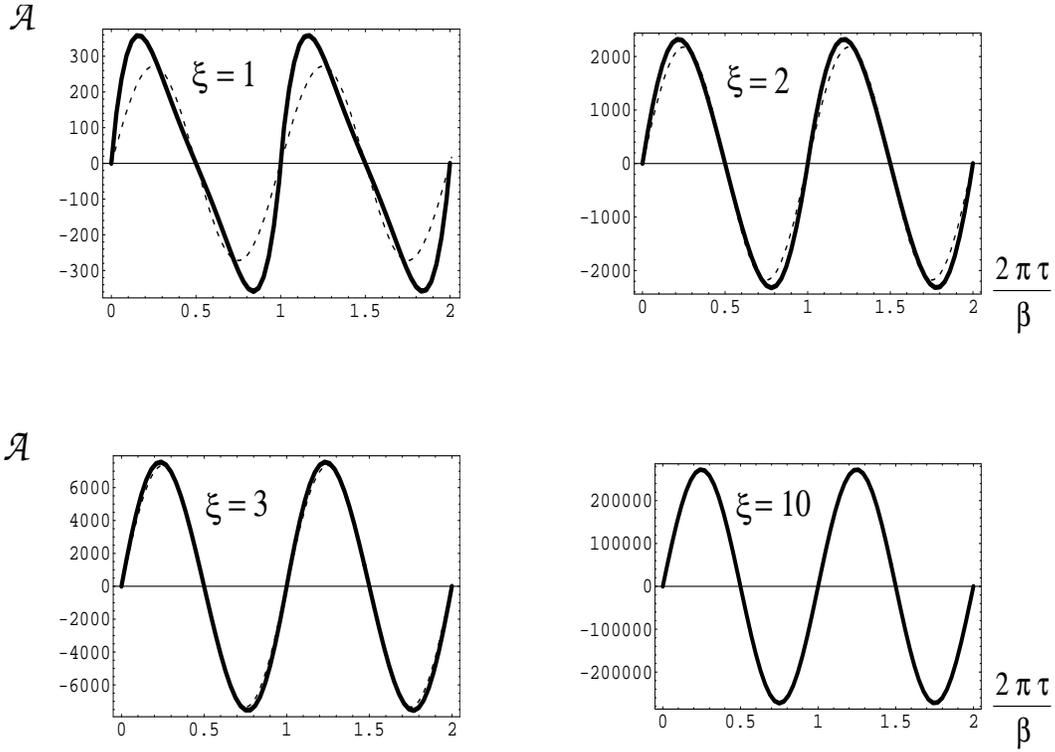}
\caption{\label{Fig-1}   								
The function $\mathcal A ( \frac{2\pi\tau}{\beta} )$ 
plotted over two periods with different values of $\xi$.
For comparison the function $272 \xi^3 \sin ( \frac{2\pi\tau}{\beta} )$ is plotted as a dashed line.} 
\end{figure}

If not stated otherwise we consider the gauge group SU(2) from now on. 
We perform a spatial coarse-graining over the 
sector of nontrivially BPS saturated (nonpropagating) field configurations in singular gauge to arrive at a 
nonpropagating adjoint scalar field $\phi$ of spacetime independent
modulus. Our strategy is 
to derive $\phi$'s second-order equation of motion and, by requiring 
compatibility with BPS saturation, to subsequently determine 
the field $\phi$ (modulus and phase) in terms 
of $T$ and a constant of integration $\Lambda$. The perturbative 
renormalizability\footnote{This is the statement that the only effect of
integrating out fluctuations in the sector with $Q=0$ 
is a resolution dependence of the gauge coupling and the 
normalization of a plane wave in a given gauge. Modulo these radiatively generated 
effects the effective action describing fluctuations in the sector with
$Q=0$ has the same form as the fundamental action. In the absence of
external sources probing the thermal system its inherent resolution
$|\phi|$ is a function of temperature. As a consequence, the effective gauge coupling is a 
function of temperature and, in the (only) physical gauge, the wave function 
normalization is the characteristic function $\chi$: $\chi$ equals unity 
if the plane wave resolves its environment by less than $|\phi|$,
and $\chi$ equals zero if the plane wave would resolve by more 
than $|\phi|$ since such a fluctuation already is integrated out.} 
of the sector with propagating gauge fields ($Q=0$) 
\cite{'tHooftVeltman-1,'tHooftVeltman-2,'tHooftVeltman-3,'tHooftVeltman-4} 
and the requirement of gauge invariance then yield a unique effective
action which is associated with a maximal resolution 
given by $\phi$'s modulus.\vspace{0.2cm}\\ 
{\sl 
If in the effective action a spatially homogeneous composite field emerges after
spatial coarse-graining over the sector of nontrivially BPS saturated field
configurations of trivial holonomy then this composite is a scalar under
rotations (O(3) scalar) and transforms in the adjoint representation of
the gauge group SU(2).} 
\begin{note}
\label{no4gauge}
The $A_4$-component and (nonlocal) products thereof are O(3) scalars only in covariant 
gauges.
\end{note}
\noindent As stated in the Introduction, the term `spatial coarse-graining' refers
to a lowering of the maximal resolution available in the system at a
given temperature when keeping the partition function fixed. The term
`effective action' refers to minus the exponent in the weight according to 
which an average over configurations is performed in the 
partition function after spatial coarse-graining.   
$\mbox{}$\\ 
{\sl exclusion of explicit nontrivial holonomy}: As shown in 
\cite{GrossPisarskiYaffe}, calorons/anticalorons with explicit
nontrivial holonomy induce a one-loop effective action which 
diverges with the three-volume of the system. Thus 
explicit holonomy must not enter the process of spatial 
coarse-graining in the caloron/anticaloron sector. 
(Short-lived implicit holonomy, however, emerges by quantum deformation
of the trivial-holonomt case and is responsible for the generation 
of short-lived magnetic dipoles (large thermodynamic weight) or screened
magnetic monopoles and antimonopoles (very small thermodynamic
weight).\\ 
{\sl transformation property under O(3)}: Since nontrivially
  BPS saturated field configurations of trivial holonomy are 
nonpropagating field configurations their coarse-grained counterparts represent spatially
homogeneous background fields in the effective theory. But the 
existence of a nontrivial O(3) tensor after coarse-graining would 
spontaneously break rotational invariance which is impossible in the
absense of microscopic degrees of freedom that single out a direction 
in space at vanishing momentum.\\   
{\sl gauge transformation property:} The scalar $\phi$ must transform 
homogeneously under a change of gauge for otherwise the 
coarse-grained gauge field $\delta A_\mu$ would have to form a composite 
to couple to $\phi$ in a gauge-invariant way. The existence of
such a composite on the level of the effective action 
would, however, contradict 
perturbative renormalizability
\cite{'tHooftVeltman-1,'tHooftVeltman-2,'tHooftVeltman-3,'tHooftVeltman-4}
which 
states that all propagating degrees of freedom are represented by 
$\delta A_\mu$ itself. The only homogeneously 
transforming, nontrivial quantities in the fundamental theory 
are (nonlocal) products of the field strength 
$F_{\mu\nu}$. Since 
\begin{equation}
\label{SU2gen}
t_at_b=\frac12\{t_a,t_b\}+\frac12[t_a,t_b]=
\frac12\left(\frac12\delta_{ab}{\mathbbm 1}_2+i\epsilon_{abc}t_c\right)
\end{equation}
we may without restriction of generality schematically write
\begin{equation*}
\phi^{{a_1}\cdots {a_K}}\sim \mbox{tr}\,\left( t^{a_1}\cdots t^{a_K}
F\cdots F\,\right)\,,\ \ \ (K\ge 1)\,
\end{equation*}
with appropriate contractions of the Lorentz indices and parallel
transports for the field strength $F\equiv F^b t_b$ implied. 
By virtue of Eq.\,(\ref{SU2gen}) this can always 
be decomposed into spin-0 and spin-1 representations 
of SU(2). The case of the spin-0 representation (gauge-invariant composite) is irrelevant because it
decouples and no energy-momentum is associated with it, see proposition \ref{emtens}. 
Thus we are left with the spin-1 representation which proves the
claim.\vspace{0.2cm}\\ 
{\sl Notation:} We denote by $\left\{(\tau,0),(\tau,\vec{x})\right\}$
the spacelike Wilson line ${\cal P}\exp\left[i\int_{(\tau,0)}^{(\tau,\vec{x})}dz_\mu A_\mu(z)\right]$ where 
the path of integration is a straight line.\vspace{.2cm}\\ 
{\sl The following is the unique definition for the set ${\cal K}$ 
of $\tau$-dependent algebra-valued functions which contains $\phi$'s 
phase $\hat{\phi}\equiv\frac{\phi}{|\phi|}\,,\ \
(|\phi|^2\equiv \mbox{tr}\,\frac12\,\phi^2)$:
\begin{equation}
\label{SU2geKdef}
{\cal K}=\Big\{\sum_{\alpha=C,A} \int d^3x\,\int d\rho\ \mbox{tr}\ \vec{t}\,
F_{\mu\nu}(\tau,\vec{0})_{\alpha}\,\left\{(\tau,\vec{0}),(\tau,\vec{x})\right\}_{\alpha}
\,F_{\mu\nu}(\tau,\vec{x})_{\alpha}\,
\left\{(\tau,\vec{x}),(\tau,\vec{0})\right\}_{\alpha}\Big\}\,,
\end{equation}
where the sum is over a HS caloron and anticaloron (singular
gauge) and $\vec{t}\equiv (t^1,t^2,t^3)$. Explicitly, the set ${\cal K}$ is parametrized by the
coordinates of the caloron/anticaloron center
$z_{C,A}=(\tau_{C,A},\vec{z}_{C,A})$. The integrals are over infinite space and the entire range
$0\le\rho\le\infty$ of the scale parameter $\rho$, and, as we shall see,
the set ${\cal K}$ is implicitly parametrized by arbitrary rescalings
and global gauge transformations.}
\vspace{0.2cm}\\  
{\sl HS caloron/anticaloron}:\\ 
Again, explicit holonomy is excluded by the result of the semiclassical
calculation in \cite{GrossPisarskiYaffe} but implicit, short-lived 
holonomy emerges by the quantum deformation of the trivial-holonomy
case. 
{\sl local definition}:\\ 
This is excluded by Proposition \ref{emtens}.\\ 
{\sl curved path for evaluation of Wilson line $\left\{(\tau,0),(\tau,\vec{x})\right\}$}:\\ 
Since the path is purely spacelike
there exists no mass scale on the level of BPS saturated field
configurations to parameterize curvature.\\  
{\sl higher $n$-point functions}:\\ 
Since ${\cal K}$ contains the
dimensionless phase $\hat{\phi}$ all its members must be
dimensionless. Considering nonlocal, 
$n$-fold products of $F_{\mu\nu}$ with $n>2$ together with the associated
additional space integrations, a factor of $\beta^{2-n}$ would have to
be introduced to make these contributions dimensionless. Since there are
no explicit dependences on $\beta$ on the
level of BPS saturated field configurations, see beginning of Sec.\,2.2, this possibility does not
exist.\\ 
{\sl moduli-space average}:\\  
(i) Integrations over the dimensionful moduli $\rho$ and $\tau_{C,A},\vec{z}_{C,A}$ 
must have a flat measure since the members of ${\cal K}$ make no reference
to any scale on the level of BPS saturated field configurations.\\ 
(ii) The right-hand side of
Eq.\,(\ref{SU2geKdef}) transforms in the adjoint representation. Shifting
the calo\-ron/anti\-caloron spatial center from $\vec{0}$ to
$\vec{z}_{C,A}$ and honoring sphe\-rical symmetry, an additional pair of
Wilson lines would have to be introduced to parallel transport
$F_{\mu\nu}$ from $(0,\vec{0})$ to $(0,\vec{z}_{C,A})$. Integrating then
over $\vec{z}_{C,A}$ yields zero. (If this integral would not vanish then the scale $\beta$
would occur explicitely in the definition of the dimensionless members
of ${\cal K}$. But this is forbidden on the level of BPS saturation,
see beginning of Sec.\,2.2.)\\   
(iii) An integration over $\tau_{C,A}$ yields zero. 
(The case of a constant term in 
the Fourier series associated with the integrand again would imply that the scale $\beta$ occurs 
explicitely in the definition of the dimensionless members
of ${\cal K}$.)\\ 
(iv) Since the members of ${\cal K}$ are gauge-variant objects
integrations 
over the global gauge orientations of the 
caloron/\-anticaloron yield zero and thus are forbidden.\\  
{\sl shift $\vec{0}\to\vec{y}\not=0$}: Shifting 
$\vec{0}\to\vec{y}$ in Eq.\,(\ref{SU2geKdef}) but leaving the caloron/\-anti\-caloron center
fixed at $\vec{0}$, spherical symmetry would imply the need for an
additional pair of Wilson lines to connect $\vec{y}$ with $\vec{0}$. 
But this is just a global gauge rotation of the unshifted situation and
thus does not alter ${\cal K}$.\\ 
{\sl caloron/anticaloron with $|Q|>1$}:\\ 
Besides the translational moduli there are $m>1$ dimensionful
moduli in such a configuration. For example, a trivial-holonomy caloron 
with $Q=2$ has three dimensionful moduli: two scale
parameters and the distance between the two centers of its topological
charge. Considering an $n$-point function ($n$ nonlocal factors of the
field strength $F_{\mu\nu}$) with $n-1$ integrations over space and a
flat-measure integration over the $m$ dimensionful moduli (not counting
the shift moduli) of the caloron, we arrive at a mass dimension
$2n-3(n-1)-m=3-n-m$ of the object. To avoid the introduction of explicit powers
of $\beta$ (BPS saturation) in the definition of ${\cal K}$ this mass
dimension needs to vanish. But for $n\ge 2$ and 
$m>1$ we have $3-n-m\not=0$.
\begin{proposition}
\label{Wline}
The Wilson line $\left\{(\tau,\vec{0}),(\tau,\vec{x})\right\}_{C,A}$
evaluates to 
\begin{equation}
\left\{(\tau,\vec{0}),(\tau,\vec{x})\right\}_{C,A}=\cos
g\pm 2i t_b\frac{x^b}{r}\,\sin g\,,
\end{equation}
where $g=g(\tau,r,\rho)=g(\beta\hat{\tau},\beta\hat{r},\beta\hat{\rho})\equiv\hat{g}(\hat{\tau},\hat{r},\hat{\rho})\equiv \int_0^1 ds\,
\frac{r}{2}\pd_{\tau}\log\Pi(\tau,sr,\rho)$. The $+$ or $-$ sign relates
to a caloron or an anticaloron, respectively. Explicitly, one has
\begin{equation}
\label{gexp}
\hat{g}=-\pi^2\hat{\rho}^2\,\sin(2\pi\hat{\tau})\int_0^1 ds\,\frac{1}{s}
\frac{\sinh(2\pi\hat{r}s)}{[\cosh(2\pi\hat{r}s)-\cos(2\pi\hat{\tau})]
[\cosh(2\pi\hat{r}s)-\cos(2\pi\hat{\tau})+\frac{\pi\hat{\rho}^2}{\hat{r}s}\sinh(2\pi\hat{r}s)]}\,.
\end{equation}
The function $\hat{g}$ exists and 
approaches constancy in $\hat{r}$ more than exponentially fast with increasing $\hat{r}>1$. 
\end{proposition}
\begin{note}
\label{phases}
To point out essential properties only we have set the phases $\tau_C$
and $\tau_A$ of the $\tau$ dependences of 
caloron and anticaloron equal to zero. They can easily be reinstated by
letting $\tau\to\tau+\tau_{C,A}$.    
\end{note}
\begin{proof}
$\mbox{}$\\ 
{\sl irrelevance of path-ordering}:\\ 
Observe that $\int_{(\tau,\vec{0})}^{(\tau,\vec{x})}\left.dz_{\mu} 
A_{\mu}(z)\right|_{C,A}=\pm\int_0^1 ds\,x_i A_i(\tau,s\vec{x})=\pm t_b
x^b\,\partial_\tau\int_0^1 ds\,\log\Pi(\tau,sr,\rho)$. That is, the
integrand in the exponent of 
$\left\{(\tau,\vec{0}),(\tau,\vec{x})\right\}_{C,A}$ varies along a
fixed direction in the Lie algebra of SU(2). Path-ordering thus can be
omitted.\\ 
{\sl explicit form of $\left\{(\tau,\vec{0}),(\tau,\vec{x})\right\}_{C,A}$ }: Routine
computation.\\ 
{\sl existence of $\hat{g}$ for all values of its arguments}:\\ 
The potentially problematic point in the domain of integration is 
$s=0$ for $\hat{\tau}=k\in {\mathbbm Z}$. By Taylor expanding the sine function
in front of the integral in Eq.\,(\ref{gexp}) about $\hat{\tau}=k$ and by
Taylor expanding the cosine and the cosine hyperbolic functions in the denominator of
the integrand about $\hat{\tau}=k$ and $s=0$, respectively, one easily checks that the limit 
$\hat{\tau}\to k$ exists when $\hat{\rho}\ge 0$ and $\hat{r}\ge 0$.\\  
{\sl saturation property for growing $\hat{r}>0$ }:\\ Split 
the integration in Eq.\,(\ref{gexp}) as ${\cal I}\equiv\int_0^1\,ds\equiv{\cal I}_1+{\cal I}_2\equiv\int_0^{\frac{1}{2\pi\hat{r}}}\,ds+\int_{\frac{1}{2\pi\hat{r}}}^1\,ds=
\int_0^1\,dz +\int_1^{2\pi\hat{r}}\,dz$. ${\cal I}_1$ does not depend on
$\hat{r}$. The integrand $I$ is given as  
\begin{equation}
I(z,\hat{\rho},\hat{\tau})\equiv \frac{\sinh z}{z[\cosh z-\cos(2\pi\hat{\tau})][\cosh
z-\cos(2\pi\hat{\tau})+\frac{2(\pi\hat{\rho})^2}{z}\,\sinh z]}\,.
\end{equation} 
For the integration in ${\cal I}_2$ the integrand $I$ is bounded from above as
\begin{equation}
\label{estimatehatr}
I(z,\hat{\rho},\hat{\tau})\le \frac{2\,e^z}{(e^z-2\cos(2\pi\hat{\tau}))^2}\,,\ \ \ \
\forall\hat{\tau},\hat{\rho}\,; z\ge 1\,.
\end{equation}
Since
$\int_1^{2\pi\hat{r}}dz\,I(z,\hat{\rho},\hat{\tau})=\int_1^{\infty}dz\,I(z,\hat{\rho},\hat{\tau})-
\int_{2\pi\hat{r}}^{\infty}dz\,I(z,\hat{\rho},\hat{\tau})$ and since, by virtue of Eq.\,(\ref{estimatehatr}), the modulus of the second summand is bounded by 
$\frac{2}{e^{2\pi\hat{r}}-2\cos(2\pi\hat{\tau})}$ we are assured a more
than exponentially fast saturation in $\hat{r}$. Numerically, 
$\frac{\int_{2\pi\hat{r}}^{\infty}dz\,I(z,\hat{\rho},\hat{\tau})}{\cal
  I}<10^{-5}$ for $\hat{r}>2$ independently of $\hat{\rho}$. (Saturation
in $\hat{\rho}$ is now trivial.) 
\end{proof}
\begin{proposition}
\label{wholeintegC}
The integrand in Eq.\,(\ref{SU2geKdef}), when evaluated on a caloron, is 
as 
\begin{equation}
\label{integrandphi}
-i\,\beta^{-2}\frac{32\pi^4}{3}\frac{x^a}{r}\frac{\pi^2\hat{\rho}^4+\hat{\rho}^2(2+\cos(2\pi\hat{\tau}))}
{\left(2\pi^2\hat{\rho}^2+1-\cos(2\pi\hat{\tau})\right)^2}\times F[\hat{g},\Pi]\,,
\end{equation}
where the functional $F$ is given as
\begin{equation}
F[\hat{g},\Pi]=2\cos(2\hat{g})\left(2\frac{[\partial_\tau\Pi][\partial_r\Pi]}{\Pi^2}-\frac{\partial_\tau\partial_r\Pi}{\Pi}\right)+
\sin(2\hat{g})\left(2\frac{[\partial_r\Pi]^2}{\Pi^2}-2\frac{[\partial_\tau\Pi]^2}{\Pi^2}+
\frac{\partial^2_\tau\Pi}{\Pi}-\frac{\partial^2_r\Pi}{\Pi}\right)\,.
\end{equation}
\end{proposition}
\begin{proof}
Lengthy routine computation, see \cite{HerbstHofmann2004,HerbstDiplom}. 
\end{proof}
\begin{proposition}
\label{wholeintegA}
The integrand in Eq.\,(\ref{SU2geKdef}), when evaluated on an
anticaloron, is obtained by a parity transformation,
$\vec{x}\to-\vec{x}$, of the integrand evaluated on a caloron. 
\end{proposition}
\begin{note}
\label{cancCA}
For equal temporal phases, $\tau_C=\tau_A$, it then follows 
that the integrands cancel. But, as we will show, nontrivial BPS saturation 
of the field $\phi$ requires that $\tau_C-\tau_A=\pm\frac{\pi}{2}$.   
\end{note}
\begin{proof}
It is easily checked that $F_{\mu\nu}(\tau,\vec{x})_C= F_{\mu\nu}(\tau,-\vec{x})_A$ and that\\ 
$\left\{(\tau,\vec{0}),(\tau,\vec{x})\right\}_C=
\left(\left\{(\tau,\vec{x}),(\tau,\vec{0})\right\}_C\right)^\dagger=\left\{(\tau,\vec{0}),(\tau,-\vec{x})\right\}_A=
\left(\left\{(\tau,-\vec{x}),(\tau,\vec{0})\right\}_A\right)^\dagger$. This proves the claim.  
\end{proof}
\begin{remark}
Due to the appearance of the factor $\frac{x^a}{r}$ in the expression
(\ref{integrandphi}) the  unconstrained angular integration in
Eq.\,(\ref{SU2geKdef}) yields zero. Thus for the final integration over
$\hat{\rho}$ to possess a nonvanishing integrand the radial integral
must diverge.   
\end{remark}
\begin{proposition}
\label{intovr}
The only term in $F[\hat{g},\Pi]$, which gives rise to the divergence of
the radial integral, is
$-\sin(2\hat{g})\frac{\partial^2_r\Pi}{\Pi}$. 
This divergence is logarithmic. 
\end{proposition}
\begin{note}
\label{BBcor}
Since only spatial derivatives are involved this term arises from
magnetic-magnetic correlations. But it is the magnetic 
sector whose insufficient screening gives rise to the poor 
convergence properties in thermal perturbation theory \cite{Linde1980}.
\end{note}
\begin{proof}
Obviously, no divergence arises when $\hat{r}\to 0$. We have
\begin{equation}
\label{expprepot1}
\partial_\tau\Pi(\tau,r)=
\beta^{-1}\partial_{\hat{\tau}}\hat{\Pi}(\hat{\tau},\hat{r})\stackrel{\hat{r}\gg
1}\longrightarrow
-\frac{(2\pi\hat{\rho})^2}{\beta\hat{r}}\sin(2\pi\hat{\tau})\exp(-2\pi\hat{r})\,,
\end{equation}
\begin{equation}
\label{expprepot2}
\partial^2_\tau\Pi(\tau,r)=
\beta^{-2}\partial^2_{\hat{\tau}}\hat{\Pi}(\hat{\tau},\hat{r})\stackrel{\hat{r}\gg
1}\longrightarrow 
\frac{2\pi}{\beta}\frac{(2\pi\hat{\rho})^2}{\beta\hat{r}}\left(4\sin(2\pi\hat{\tau})\exp(-4\pi\hat{r})-
\cos(2\pi\hat{\tau})\exp(-2\pi\hat{r})\right)\,. 
\end{equation}
Thus all terms in $F[\hat{g},\Pi]$ containing $\partial_{\tau}\Pi$ or $\partial^2_{\tau}\Pi$
give rise to finite contributions to the radial integral. (The measure
is $d\hat{r}\,\hat{r}^2$.) The same
holds true for the term with $(\partial_{r}\Pi)^2$ since 
\begin{equation}
\label{expprepot3}
\Pi(\tau,r)\equiv\hat{\Pi}(\hat{\tau},\hat{r})\stackrel{\hat{r}\gg 1}
\longrightarrow 1+\frac{\pi\hat{\rho}^2}{\hat{r}}\Rightarrow (\partial_r\Pi(\tau,r))^2\stackrel{\hat{r}\gg 1}
\longrightarrow \beta^{-2}\frac{\pi^2\hat{\rho}^4}{\hat{r}^4}\,.
\end{equation}
But 
\begin{equation}
\label{expprepot4}
\partial^2_r\Pi(\tau,r))\stackrel{\hat{r}\gg 1}
\longrightarrow \beta^{-2}\frac{2\pi\hat{\rho}^2}{\hat{r}^3}\,.
\end{equation}
Thus 
\begin{equation}
\label{expprepotconc}
-\int_0^{\infty}dr\,r^2\,\sin(2g)\frac{\partial^2_r\Pi}{\Pi}\sim 
-\beta\left(\mbox{finite}+2\pi\hat{\rho}^2\left(\lim_{\hat{r}\to\infty}
    \sin(2\hat{g})\right)\int_{\hat{R}}^\infty \frac{d\hat{r}}{\hat{r}}\right)\,,
\end{equation}
where the $\sim$ sign indicates 
that the right-hand side approaches the left-hand side more than 
exponentially fast for increasing $\hat{R}>1$, see Proposition
\ref{Wline}. Obviously, the integral in
Eq.\,(\ref{expprepotconc}) diverges logarithmically. 
\end{proof}
\begin{remark}
\label{sumtobeev}
In summary, the quantity to be evaluated is
\begin{equation}
\label{sumeval}
i\frac{64\pi^5}{3}\int d\hat{\rho}\,\hat{\rho}^2\,
\frac{\pi^2\hat{\rho}^4+\hat{\rho}^2(2+\cos(2\pi\hat{\tau}))}
{\left(2\pi^2\hat{\rho}^2+1-\cos(2\pi\hat{\tau})\right)^2}\int d\Omega\frac{x^a}{r}\int_{\hat{R}}^\infty
\frac{d\hat{r}}{\hat{r}}\sin(2\hat{g})\,.
\end{equation}
\end{remark}
\begin{remark}
\label{angreg}gauge-invariant way
The angular integration in the expression (\ref{sumeval}),
\begin{equation}
\label{angintreg} 
\int
d\Omega\frac{x^a}{r}=\int_{-1}^{+1}d(\cos\theta)\int_{\alpha_C}^{\alpha_C+2\pi}d\varphi\,\frac{x^a}{r}\,,\
\ \ (0\le\alpha_C\le 2\pi)\,,
\end{equation} 
is regularized by introducing a
defect/surplus angle $\eta^\prime\ll 1$ for
the azimuthal integration in $\varphi$: $\alpha_C\to\alpha_C\pm\eta^\prime$
(lower integration limit) and  $\alpha_C\to\alpha_C\mp\eta^\prime$
(upper integration limit). This singles out a unit vector 
$\hat{n}_C\equiv(\cos\alpha_C,\sin\alpha_C,0)$. Obviously, 
a rotation of $\hat{n}_C$ is induced by a rotation of the 
cartesian coordinates in which the transition to polar 
coordinates is performed. But for $\hat{\phi}\in {\cal K}$ this 
amounts to nothing but a global gauge 
rotation. Thus no breaking of rotational symmetry is 
introduced by the angular regularization.  
\end{remark}
\begin{proposition}
\label{rhointres}
Without restriction of generality the contribution from the 
anticaloron is also regularized in the $x_1x_2$-plane with angle
$\alpha_A$ (global gauge choice). Then we arrive at 
\begin{equation}
\label{calantical}
{\cal K}=\Xi_C(\delta^{a1}\cos\alpha_C+\delta^{a2}\sin\alpha_C)\,{\cal
  A}\left(2\pi(\hat{\tau}+\hat{\tau}_C)\right)+\Xi_A(\delta^{a1}\cos\alpha_A+\delta^{a2}\sin\alpha_A)\,{\cal
  A}\left(2\pi(\hat{\tau}+\hat{\tau}_A)\right)\,,
\end{equation}
where $\Xi_C,\Xi_A\in {\mathbbm R}$ (undetermined: 0 [angular
integr.]$\times\infty$ 
[radial integr. subject to dimensional smearing, see \cite{HerbstHofmann2004}]) 
and $0\le\hat{\tau}_C,\hat{\tau}_A\le 1$ (undetermined:
modulus of caloron/anticaloron which cannot be averaged over, see
argument (iii) above Prop. 2.). The function ${\cal
  A}(2\pi\hat{\tau})$ in Eq.\,(\ref{calantical}) is given as 
\begin{equation}
\label{calA}
{\cal A}(2\pi\hat{\tau})=\frac{32\pi^7}{3}\int_0^{\xi} d\hat{\rho}\,\hat{\rho}^4\,
\left[\lim_{\hat{r}\to\infty}\sin(2\hat{g}(\hat{\tau},\hat{r},\hat{\rho}))\right]\frac{\pi^2\hat{\rho}^2+
\cos(2\pi\hat{\tau})+2}{\left(2\pi^2\hat{\rho}^2-\cos(2\pi\hat{\tau})+1\right)^2}\,.
\end{equation}
The integral over $\hat{\rho}$ in Eq.\,(\ref{calA}) diverges cubically
for $\xi\to\infty$.
\end{proposition}
\begin{proof}     
Routine computation using the fact that $\hat{g}$ saturates for
$\hat{\rho}\to\infty$ (see Prop.\,\ref{Wline}).
\end{proof}   
\begin{theorem}
\label{sinef}
The function ${\cal A}(2\pi\hat{\tau})$ rapidly approaches
$\mbox{const}_\infty\times\xi^3\sin(2\pi\hat{\tau})$ where $\mbox{const}_\infty=272.018$.  
\end{theorem}
\begin{proof} 
Since $\hat{g}$ saturates one may evaluate the integral numerically thus
proving the claim.  
\end{proof} 
\begin{remark}
\label{rhosat}
Already for $\xi=3$ one has
$\frac{\mbox{const}_{3}-\mbox{const}_\infty}{\mbox{const}_\infty}=0.025$,and
the functional dependence on $\hat{\tau}$ practically is a sine, see
Fig.\,\ref{Fig-1}. Since there is such a fast saturation towards a sine 
function the prefactor $272.018\times\xi^3$, which is computed
numerically, 
can be absorped into the
undetermined, real number  $\Xi_{C,A}$ in Eq.\,(\ref{calantical}). In this
sense the result for ${\cal K}$ is independent of the cutoff $\xi$ for 
$\xi$ sufficiently large, see again Fig.\,\ref{Fig-1}.  
\end{remark}
\begin{theorem}
\label{diffop}
The set ${\cal K}$ coincides with the kernel of the linear differential 
operator ${\cal
  D}\equiv\partial^2_\tau+\left(\frac{2\pi}{\beta}\right)^2$ acting on
an adjoint scalar field $\hat{\phi}$ with two polarizations. Thus
${\cal D}$ is uniquely determined by ${\cal K}$.   
\end{theorem}
\begin{proof} 
There are two independent `polarizations' contained in ${\cal K}$ which
are given by the unit vectors $\hat{n}_C$ and $\hat{n}_A$. For each polarization there is an undetermined phase
shift $\hat{\tau}_{C,A}$ and an undetermined amplitude
$|\Xi_{C,A}|$, and each polarization is annihilated by ${\cal
  D}$. Modulo global gauge rotations there are thus 
two real parameters for each polarization of $\hat{\phi}$ which span the
solution space of ${\cal D}\hat{\phi}=0$, and ${\cal D}$ is determined
uniquely.    
\end{proof} 
$\mbox{}$\vspace{0.2cm}\\ 
${\cal D}$ is linear. Under the ansatz that $|\phi|$ is spacetime
independent\footnote{Compare with discussion in the Introduction: The 
search for configurations $\phi$ with constant modulus and, up to global
gauge transformations, pre-determined phase $\hat{\phi}$ 
leads to the existence of a unique and consistent (no contradiction to saturation 
of $\hat{\phi}$) value of $|\phi|$ and thus to the according maximal 
resolution.} Thm.\,\ref{diffop} implies that the field
$\phi$ possesses a canonic kinetic term 
$\mbox{tr}\,(\partial_\tau \phi)^2$ in its effective, euclidean 
Langrangian density.
\begin{theorem}
\label{lambdaemrg}
The adjoint scalar field $\phi$ is subject to the euclidean 
Lagrangian density 
\begin{equation}
\label{effactphi}
{\cal
  L}_\phi=\mbox{tr}\left((\partial_\tau\phi)^2+V(\phi^2)\right)
\end{equation} 
with $V(\phi^2)\equiv\Lambda^6\phi^{-2}$ and $\Lambda$ an arbitrary 
mass scale. Here $\phi^{-1}\equiv \frac{\phi}{|\phi|^2}$. 
\end{theorem}
\begin{proof} 
Due to the BPS saturation of coarse-grained calorons/anticalorons no
explicit temperature dependence may appear in $\phi$'s 
effective action. Together with the statement preceding
Thm.\,\ref{lambdaemrg} this implies an
effective action of the form (\ref{effactphi}) with a yet unknown 
potential $V(\phi^2)$. The according Euler-Lagrange 
equations\footnote{Notice our notational convention: $V$ is either a scalar-valued function of its
scalar-valued argument or a matrix-valued function of its matrix-valued
argument. In both cases the functional dependence is identical.} 
are \cite{GH2006}
\begin{equation}
\label{ELe}
\partial^2_\tau\phi^a=\frac{\partial V(|\phi|^2)}{\partial
  |\phi|^2}\,\phi^a\ \ (\mbox{in components})\ \ \Leftrightarrow\ \ \partial^2_\tau\phi=\frac{\partial V(\phi^2)}{\partial
  \phi^2}\,\phi\ \ (\mbox{in matrix form})\,.
\end{equation}  
Since $\phi$'s motion is within a plane in the three-dimensional vector
space of the SU(2) Lie algebra, since $|\phi|$ is independent of
spacetime, and since $\phi$'s phase $\hat{\phi}$ is of 
period unity, see Thm.\,\ref{diffop}, one may, without restriction of generality (global gauge
choice), write the solution to Eq.\,(\ref{ELe}) as
\begin{equation}
\label{solEL}
\phi=2\,|\phi|\,t_1\,\exp(\pm\frac{4\pi i}{\beta}t_3\tau)\,.
\end{equation} 
BPS saturation, or equivalently, the vanishing of the euclidean energy
density and Eq.\,(\ref{solEL}) imply
\begin{equation}
\label{BPSans}
|\phi|^2\left(\frac{2\pi}{\beta}\right)^2-V(|\phi|^2)=0\,.
\end{equation}
On the other hand, comparing
$\partial^2_\tau\phi+\left(\frac{2\pi}{\beta}\right)^2\phi=0$, see
Thm.\,\ref{diffop}, 
with Eq.\,(\ref{ELe}), we have
\begin{equation}
\label{ELcomp}
\left(\frac{2\pi}{\beta}\right)^2=-\frac{\partial V(|\phi|^2)}{\partial
  |\phi|^2}\,.
\end{equation}
Together, Eqs.\,(\ref{BPSans}) and (\ref{ELcomp}) yield
\begin{equation}
\label{eomPot}
\frac{\partial V(|\phi|^2)}{\partial
  |\phi|^2}=-\frac{V(|\phi|^2)}{|\phi|^2}\,.
\end{equation}
Eq.\,(\ref{eomPot}) is a first-order differential 
equation whose solution is
\begin{equation}
\label{solPot} 
V(|\phi|^2)=\frac{\Lambda^6}{|\phi|^2}\,,
\end{equation}
where $\Lambda$ denotes an arbitrary mass scale (the Yang-Mills scale).
\end{proof} 
\begin{corollary}
\label{phimod}
The modulus of the field $\phi$ is given as
$|\phi|=\sqrt{\frac{\Lambda^3\beta}{2\pi}}$ and hence, modulo a global
change of gauge, 
\begin{equation}
\label{phirel}
\phi=2\,\sqrt{\frac{\Lambda^3\beta}{2\pi}}\,t_1\,\exp(\pm\frac{4\pi
  i}{\beta}t_3\tau)\,.
\end{equation}
\end{corollary}
\begin{proof} 
Substitute Eq.\,(\ref{solPot}) into Eq.\,(\ref{BPSans}), solve for
$|\phi|$ and substitute the result into Eq.\,(\ref{solEL}). 
\end{proof}
\begin{remark}
\label{inertness}
The field $\phi$ represents coarse-grained nonpropagating,
noninteracting, 
BPS saturated field configurations of topological charge modulus $|Q|=1$ 
and trivial holonomy. Thus it should itself not propagate. This is 
explicitly checked by computing the mass $M_{\delta\phi}$ 
of potential fluctuations $\delta\phi$ about the configuration in 
Eq.\,(\ref{phirel}) as
\begin{equation}
\label{phimass}
M_{\delta\phi}^2=\left.2\frac{\partial^2V}{\partial|\phi|^2}\right|_{|\phi|=\sqrt{\frac{\Lambda^3\beta}{2\pi}}}=48\pi^2T^2=12\lambda^3|\phi|^2\,,
\end{equation}
where $\lambda\equiv\frac{2\pi T}{\Lambda}$. Since $\lambda$ is
considerably larger than unity in the deconfining phase, see below 
where it is derived that $\lambda\ge\lambda_c=13.87$, and since
the scale $|\phi|$ represents the maximal resolution 
(off-shellness) of any fluctuation after coarse-graining\footnote{Notice
that only the case of pure thermodynamics is discussed here. If an
external probe is applied to the thermal system then an 
additional momentum scale enters, and $|\phi|$ no longer represents the
only scale of resolution.} 
we conclude that the field $\phi$ does not fluctuate:
neither thermally nor quantum mechanically. The field $\phi$ thus
represents a spatially homogeneous background for the dynamics of the 
coarse-grained, propagating gauge field (sector with $Q=0$).   
\end{remark}
Notice that with $\lambda_c=13.87$ one obtains $\frac{|\phi|^{-1}}{\beta}\ge
8.221\times\left(\frac{\lambda}{\lambda_c}\right)^{3/2}\,,\ (\lambda\ge\lambda_c)$. But for 
$\hat{r}=8.221\times\left(\frac{\lambda}{\lambda_c}\right)^{3/2}$ the
exponentially suppressed term below 
Eq.\,(\ref{estimatehatr}) is a correction of less than one in 
$10^{22}$! At the same time, setting 
$\xi=8.221\times\left(\frac{\lambda}{\lambda_c}\right)^{3/2}$ in 
Eq.\,(\ref{calA}), one is deep inside the saturation regime for 
the set ${\cal K}$, see Fig.\,\ref{Fig-1}. Thus, with a maximal
resolution 
$|\phi|$ in the effective theory (corresponding to a length scale $|\phi|^{-1}$ up to 
which short-distance fluctuations in the fundamental fields are coarse-grained over to derive the
effective theory) the infinite-volume limit used to derive ${\cal K}$ and in
turn the differential operator ${\cal D}$ is extremely well
approximated.      

According to Rem.\,\ref{inertness} the configuration in 
Eq.\,(\ref{phirel}) is not altered by interactions with 
the gauge fields in the sector with $Q=0$. Thus the nonperturbative emergence of the
scale $\Lambda$ is not influenced by this sector. Compare this with the 
situation in perturbation theory at $T=0$ where $\Lambda$ is the pole
position for the evolution of the fundamental gauge 
coupling $g$ in the sector with $Q=0$. There, the value of $\Lambda$ is 
dictated by the value of $g$ at a given resolution, and two 
assumptions enter. First, one assumes  properties of the perturbative
expansion that are sufficiently close to those of an asymptotic 
series to justify the low-order truncation of the beta function. 
Second, one assumes that close to the pole the perturbative prediction
for the evolution of $g$ can smoothly be extrapolated to the regime
where $g\gg 1$. For an extended discussion see \cite{GH2007beta}.       

For the case of SU(3) the field $\phi$ winds in each of the three
(nearly independent) 
SU(2) subalgebras for a third of the period $\beta$, for details see 
\cite{Hofmann2005L}.\vspace{0.2cm}\\ 
{\sl After spatial coarse-graining the effective 
Lagrangian density, subject to a maximal resolution $|\phi|$ for
propagating gauge fields, is given as 
\begin{equation}
\label{fullactden}
{\cal L}_{\tiny\mbox{eff}}[a_\mu]=\mbox{tr}\,\left(\frac12
  G_{\mu\nu}G^{\mu\nu}+(D_\mu\phi)^2+\frac{\Lambda^6}{\phi^2}\right)\,,
\end{equation}
where $G_{\mu\nu}=\partial_\mu a_\nu-\partial_\nu
a_\mu-ie[a_\mu,a_\nu]\equiv G^a_{\mu\nu}\,t_a$, $a_\mu=a_\mu^a\,t_a$ is
the coarse-grained, propagating gauge field in the sector with $Q=0$, 
$D_\mu\phi=\partial_\mu\phi-ie[a_\mu,\phi]$, and $e$ is the effective
gauge coupling.}\vspace{0.2cm}\\  
Why is this statment true? The form of the term $\frac12\,G^2$ is as in the fundamental
theory\footnote{Notice that in contrast to Eq.\,(\ref{fundpartfunct}) 
we have not absorped the coupling into the gauge field. A gauge
transformation acting on $\phi$ and $a_\mu$ thus reads: $\phi\to
\Omega\phi\Omega^\dagger$ and $a_\mu\to\Omega
a_\mu\Omega^\dagger+\frac{i}{e}\Omega\partial_\mu\Omega^\dagger\,,\ \ \Omega\in\mbox{fund(SU(2))\,or\,fund(SU(3))}$.} due to perturbative renormalizability 
\cite{'tHooftVeltman-1,'tHooftVeltman-2,'tHooftVeltman-3,'tHooftVeltman-4}, and
the only gauge-invariant way to couple the Lagrangian density of
Eq.\,(\ref{effactphi}) to the coarse-grained sector with $Q=0$, which
itself cannot generate any composite of the field $a_\mu$, is to do
the replacement $\partial_\tau\phi\rightarrow D_\mu\phi$. When we say `the only gauge-invariant way'  
we only consider local effective vertices of $\phi$ with $a_\mu$ which do not involve three or more external legs of the latter. 
(Nonlocal interactions always can be expanded into a local series involving powers of covariant derivatives.)
Such vertices would mediate energy-momentum exchange from the sector Q=0 to $|Q|=1$ {\sl after} 
coarse-graining. This, however, would contradict the BPS nature of the field $\phi$ and thus its very existence. But $\phi$'s 
existence has just been established. Therefore no vertices involving the field $\phi$ and $a_\mu$ other than the mass operator 
contained in tr\,$(D_\mu\phi)^2$ (unitary gauge, no energy-momentum transfer but massiveness of off-Cartan modes after 
infinite resummation of mass insertion) exist.

The fluctuating field $a_\mu$ is integrated out loop expanding the
logarithm of the partition function 
about the free quasiparticle situation. This loop expansion 
is nontrivial due to the term $ie[a_\mu,a_\nu]$ in $G_{\mu\nu}$ which
leads to the occurence of three-vertices and four-vertices. The momentum
transfer in these vertices is subject 
to constraints imposed by the existence of the maximal resolution
$|\phi|$. The evolution of the {\sl effective} coupling $e$ 
is determined by the invariance of Legendre transformations between thermodynamic
quantities under the applied coarse-graining up to a given loop order,
see below.\vspace{0.2cm}\\ 
{\sl Apart from (small) radiative corrections  and
modulo global gauge transformations the full ground state of the effective theory in the deconfining phase (taking into account the
interactions between and fundamental radiative modifications of 
calorons and anticalorons) is given by the configuration
in Eq.\,(\ref{phirel}) and the pure-gauge configuration
$a^{\tiny\mbox{gs}}_\mu=\mp\delta_{\mu 4}\frac{2\pi}{e\beta}\,t_3$.}
\vspace{0.2cm}\\  
The following Euler-Lagrange equation for $a_\mu$ is implied by ${\cal
  L}_{\tiny\mbox{eff}}$ in Eq.\,(\ref{fullactden}):
\begin{equation}
\label{ELgsa}
D_{\mu}G^{\mu\nu}=ie[\phi,D^\nu\phi]\,.
\end{equation}
But Eq.\,(\ref{ELgsa}) is solved by $\phi$ and
$a^{\tiny\mbox{gs}}_\mu$ by virtue of
$G_{\mu\nu}[a^{\tiny\mbox{gs}}_\kappa]=D^\nu[a^{\tiny\mbox{gs}}_\kappa]\phi\equiv
0$.

Due to $G_{\mu\nu}[a^{\tiny\mbox{gs}}_\kappa]=D^\nu[a^{\tiny\mbox{gs}}_\kappa]\phi\equiv
0$ the ground-state associated Lagrangian density is given as ${\cal
  L}_{\tiny\mbox{eff}}[a^{\tiny\mbox{gs}}_\mu]=\mbox{tr}\,\frac{\Lambda^6}{\phi^2}=4\pi\Lambda^3\,T\,,\ \ (T\equiv\beta^{-1})$. 
Thus interactions between and radiative corrections within calorons and 
anticalorons lift the energy density $\rho^{\tiny\mbox{gs}}$ 
of the ground state from zero in the case of BPS saturation to 
$\rho^{\tiny\mbox{gs}}=4\pi\Lambda^3\,T$. The fact that the ground-state
pressure 
$P^{\tiny\mbox{gs}}$ is negative,
$P^{\tiny\mbox{gs}}=-\rho^{\tiny\mbox{gs}}$, is microscopically 
explained by small-holonomy calorons and anticalorons having their BPS 
magnetic monopoles-antimonopole constituents
\cite{NahmP,NahmL,LeeLu1998,KraanVanBaalNPB1998-1,KraanVanBaalNPB1998-2,KraanVanBaalNPB1998-3} 
attract one another under
the influence of radiative corrections, for a detailed discussion which
is based on the work in \cite{Diakonov2004}, see
\cite{Hofmann2005L,Hofmann2006}. The case of the `excitation' of a 
large holonomy, which according to \cite{Diakonov2004}
leads to the dissociation of the associated 
caloron/anticaloron and hence to the liberation of a screened magnetic
monopole and its antimonopole, is extremely rare
\cite{Hofmann2005L}. Deviations from the equation of state 
$P^{\tiny\mbox{gs}}=-\rho^{\tiny\mbox{gs}}$, which are due to those 
nonrelativistic, screened magnetic monopoles and 
antimonopoles, are described in part by the radiative corrections to
the total pressure and energy density in the effective theory.       
\begin{proposition}
\label{adunitgauge}
In the effective theory the winding gauge, where $\phi$ is as in Eq.\,(\ref{phirel}) and
$a_\mu^{\tiny\mbox{gs}}$ is as above, and the unitary
gauge, where $\phi=2\,|\phi|\,t_3$, $a_\mu^{\tiny\mbox{gs}}=0$, are connected by a singular but 
admissible periodic gauge transformation. Under this gauge transformation the Polyakov 
loop $P[a_\mu^{\tiny\mbox{bg}}]$ is transformed from $P=-{\mathbbm 1}$
to $P={\mathbbm 1}$ which points out the electric ${\mathbbm Z}_2$
degeneracy of the ground state and thus deconfinement.
\end{proposition}
\begin{proof}
Since $\phi\rightarrow
{\tilde\Omega}(\tau)\phi{\tilde\Omega}^\dagger(\tau)$ under the gauge
transformation it is easily checked that $\tilde\Omega(\tau)$ is given as
\begin{equation}
\label{omegawu}
\tilde\Omega(\tau)=\Omega_{\tiny\mbox{gl}}\,Z(\tau)\,\Omega(\tau)\,,
\end{equation}
where $\Omega(\tau)\equiv\exp[\pm 2\pi i\frac{\tau}{\beta}t_3]$, 
$Z(\tau)=\left(2\Theta(\tau-\frac{\beta}{2})-1\right){\mathbbm 1}_2$, and
$\Omega_{\tiny\mbox{gl}}=\exp[i\frac{\pi}{2}t_2]$. The function $\Theta$
is defined as
\begin{equation}
\label{thetastep}\Theta(x)=\left\{\begin{array}{c}
0\,,\ \ \ \ (x<0)\,,\\ 
\frac{1}{2}\,,\ \ \ \ (x=0)\,,\\ 
1\,,\ \ \ \ (x>0)\,.
\end{array}\right.\,
\end{equation}
Thus $\tilde\Omega(\tau)$ is periodic but not smooth. The periodicity of
fluctuations $\delta a_\mu$, however, is not affected by this
gauge transformation. Namely, writing
$a_\mu=a^{\tiny\mbox{bg}}_\mu+\delta a_\mu$, we have
\begin{equation}
\label{omegatilde1}
a_\mu\rightarrow\tilde{\Omega}(a^{\tiny\mbox{gs}}_\mu+\delta a_\mu)\tilde{\Omega}^\dagger+
\frac{i}{e}\tilde{\Omega}\pd_\mu\tilde{\Omega}^\dagger 
=\Omega_{\tiny\mbox{gl}}\left(\Omega(a^{\tiny\mbox{gs}}_\mu+\delta a_\mu)\Omega^\dagger+
\frac{i}{e}\left(\Omega\pd_\mu\Omega^\dagger+Z\pd_\mu Z\right)\right)\Omega^\dagger_{\tiny\mbox{gl}}\nonumber 
\end{equation}
\begin{equation}
\label{omegatilde2}
\hspace{.5cm}=\Omega_{\tiny\mbox{gl}}\left(\Omega\delta a_\mu\Omega^\dagger+
\frac{2i}{e}\delta(\tau-\frac{\beta}{2})Z\right)\Omega_{\tiny\mbox{gl}}^\dagger
=\Omega_{\tiny\mbox{gl}}\Omega\,\delta
a_\mu\,(\Omega_{\tiny\mbox{gl}}\Omega)^\dagger\,.
\end{equation}
Now $\Omega_{\tiny\mbox{gl}}\Omega(\tau=0)=-\Omega_{\tiny\mbox{gl}}\Omega(\tau=\beta)$.
Thus the periodicity of the fluctuation $\delta a_\mu$ is unaffected by
the gauge transformation induced by $\tilde\Omega(\tau)$ (admissibility
of this change of gauge). To show the claimed transformation of the Polyakov 
loop on $a_\mu^{\tiny\mbox{bg}}$ is trivial.    
\end{proof}
One can easily show that for SU(3) the Polyakov loop 
$P[a_\mu^{\tiny\mbox{bg}}]$ forms a three-dimensional representation 
of the electric center symmetry ${\mathbbm Z}_3$, see \cite{Hofmann2005L}. 
Thus also for SU(3) the deconfining property of the thermal ground 
state follows.  

\subsection{Thermal quasiparticle excitations}

In this section we obtain the tree-level mass spectrum for emergent 
thermal quasiparticles in the effective theory, and 
we derive the evolution of the effective gauge coupling 
$e$. Next, we give analytic expressions for the temperature dependence
of thermodynamic quantities on the level of free quasiparticle 
fluctuations. We also comment on the trace anomaly of the
energy-momentum tensor.
\vspace{0.2cm}\\ 
{\sl We refer to an excitation, which possesses a temperature-dependent 
mass on tree-level in the effective theory, as a thermal quasiparticle.}
\begin{proposition}
\label{masses}
In the effective theory dynamical 
gauge symmetry breaking SU(2)$\to$U(1) is manifested for the sector with 
$Q=0$ by quasiparticle masses $m_a$. One has 
\begin{equation}
\label{massform}
m^2_a=-2e^2\mbox{tr}\,[\phi,t_a][\phi,t_a]\,.
\end{equation}
Thus, $m^2\equiv m_1^2=m_2^2=4e^2\frac{\Lambda^3}{2\pi T}$ and $m_3=0$. 
\end{proposition}
\begin{proof}
Since $a^{\tiny\mbox{gs}}_\mu=0$ in unitary gauge formula
(\ref{massform}) can be read off from the Lagrangian density ${\cal
  L}_{\tiny\mbox{eff}}$ in Eq.\,(\ref{fullactden}), and since
$\phi=2\,|\phi|\,t_3$ in unitary gauge the explicit expression for the
mass $m$ follows.  
\end{proof}
\begin{remark}
\label{complfixedgauge}
Imposing unitary gauge, with gauge condition $\phi=2\,|\phi|\,t_3$,
$a_\mu^{\tiny\mbox{gs}}=0$, and in addition Coulomb gauge for the
unbroken U(1) subgroup, with gauge condition $\partial_i a^3_i=0$,
yields a completely fixed gauge if the real-valued gauge function
$\theta$ in $\Omega_3\equiv\exp(i\theta t_3)$ 
vanishes at spatial infinity. This gauge is physical because it exhibits the quasiparticle mass 
spectrum, the physical number of polarizations -- three for $a=1,2$ and
two for $a=3$ --, and the transversality of the gauge field $a^3_\mu$ 
associated with the unbroken subgroup U(1).      
\end{remark}
\begin{remark}
\label{complfixedgaugeSU(3)}
For SU(3) the unbroken subgroup is U(1)$^2$ and six out of 
eight independent directions in the SU(3) Lie algebra acquire mass, for details
see \cite{Hofmann2005L,Hofmann2005S}.  
\end{remark}
Notice that the number of degrees of freedom before coarse-graining
matches those after coarse graining. Namely, for SU(2) one has three species of
propagating gauge fields ($Q=0$ sector) times two polarizations each plus
two species of charge-one scalar magnetic monopoles ($|Q|=1$
sector\footnote{According to the definition of the set ${\cal K}$ in
  Sec.\,3.1 topologically
  nontrivial field configurations with $|Q|=1$ only contribute to the thermal
  ground state. As a consequence, only magnetic monopoles of magnetic charge
  modulus unity occur as their constituents.}) before coarse-graining and two
species of massive gauge fields times three polarization each plus one 
species of massless gauge field times two polarizations each. In both
cases one obtains eight degrees of freedom. For SU(3) one obtains 22 
degrees of freedom before and after coarse-graining. 

In unitary-Coulomb gauge and on the level of free quasiparticles the
real-time propagators
\footnote{An analytic continuation $\tau\rightarrow -it\,, t\ \mbox{and}\ \tau\
\mbox{real}\,,$ is performed, see \cite{Kapusta}.} of the fields 
$a_\mu^{1,2}$ and $a_\mu^3$ are given as
\begin{equation}
\label{prop12}
D^{\tiny\mbox{1 or 2}}_{\mu\nu}(p)=-\tilde{D}_{\mu\nu}
\left[\frac{i}{p^2-m^2+i0}+2\pi\delta(p^2-m^2)n_B(|p_0|/T) \right]\,,
\end{equation}
\begin{equation}
\label{prop3}
D^{\tiny\mbox{3}}_{\mu\nu}(p)=-
\left\lbrace P^T_{\mu\nu}\left[\frac{i}{p^2+i0}+2\pi\delta(p^2)n_B(|p_0|/T)\right]
-i\frac{u_\mu u_\nu}{\vec{p}^2}\right\rbrace \,,
\end{equation}
where $\tilde{D}_{\mu\nu}=\left( g_{\mu\nu}-\frac{p_\mu 
p_\nu}{m^2}\right)$, $P^{00}_T = P^{0i}_T=P^{i0}_T = 0\,,\ 
P^{ij}_T=\delta^{ij}-p^{i}p^{j}/\vec{p}^2$, $u=(1,0,0,0)$
represents the four-velocity of the heat bath, and $n_B(x)=1/(e^x-1)$ denotes the 
Bose-Einstein distribution function. 

Because of the existence of a maximal resolution scale $|\phi|$ 
the deviation of the momentum $p_\mu$ in
Eqs.\,(\ref{prop12},\ref{prop3}) from its mass shell is 
constrained as
\begin{equation}
\label{cc1}
|p^2|\le|\phi|^2\,,\ \ (\mbox{for $a$=3})\,,\ \ |p^2-m^2|\le|\phi|^2\,,\ \
(\mbox{for $a$=1,2})\,.
\end{equation}
Conditions (\ref{cc1}) fix the momentum transfer in a 
three-vertex by momentum conservation. 

The following conditions fix the momentum tranfer in a four-vertex:
\label{compconst2}
\begin{equation}
\label{cc2}
|(p_1+p_2)^2|\le|\phi|^2\,,\ \ (s\ \mbox{channel})\,;\ \ 
|(p_3-p_1)^2|\le|\phi|^2\,,\ \ (t\ \mbox{channel})\,;\ \ 
|(p_2-p_3)^2|\le|\phi|^2\,,\ \ (u\ \mbox{channel})\,.
\end{equation}
These conditions\footnote{In calculating radiative 
corrections we have checked that under $|\phi|\longrightarrow
\xi|\phi|$ in (\ref{cc1}) and (\ref{cc2}), 
where $\xi$ is of order unity, the results are remarkably stable.} follow from the fact that 
massless intermediate modes in the fundamental 
theory, which do not exist in the effective theory but dress the four-vertex such that the latter appears to be
local, may not resolve distances smaller than 
$|\phi|^{-1}$ \cite{Hofmann2006}.\vspace{0.2cm}\\ 
{\sl On the one-loop level in the effective theory (gas of noninteracting
thermal quasiparticles and massless excitations) the
contribution $\Delta V$ of quantum fluctuations (arising from terms without the
factor $n_B$ in Eqs.\,(\ref{prop12},\ref{prop3})) is negligibly
small.}\vspace{0.2cm}\\ 
We estimate $\Delta V$ by the 
contribution of the massless mode $a^3_\mu$ appropriately weighted by
the number of polarizations of all fields $a^1_\mu, a^2_\mu,$ and $a^3_\mu$:
\begin{equation}
\label{delV} 
|\Delta
V|\le\frac{1}{\pi^2}\int_0^{|\phi|}dp\,p^3\,\log\left(\frac{p}{|\phi|}\right)=\frac{\phi^4}{16\pi^2}=
\frac{\lambda^{-3}}{32\pi^2}\,V\,.
\end{equation}
Since $\lambda$ is considerably larger than unity $\Delta V$ can safely be
neglected. 
\begin{theorem}
On the one-loop level the evolution of the effective coupling $e$ is determined by the
following first-order ordinary differential equation:
\begin{equation}
\label{evoleqsu2}
\partial_a\lambda=-\frac{24\lambda^4
  a}{(2\pi)^6}\frac{D(2a)}{1+\frac{24\lambda^3a^2}{(2\pi)^6}D(2a)}\,,
\end{equation}
where $a\equiv\frac{m}{2T}=2\pi e\lambda^{-3/2}$ and $D(y)\equiv\int_0^\infty
dx\,\frac{x^2}{\sqrt{x^2+y^2}}\frac{1}{\exp(\sqrt{x^2+y^2})-1}$. The
evolution governed by Eq.\,(\ref{evoleqsu2}) possesses two 
fixed points: $a=0$ and $a=\infty$. The latter is associated with 
a critical temperature $\lambda_c$ of value $\lambda_c=13.87$. An attractor to the 
evolution exists. It is given as
$a(\lambda)=4\sqrt{2}\pi^2\lambda^{-3/2}$ for $\lambda\gg\lambda_c$ and
$a(\lambda)\propto -\log(\lambda-\lambda_c)$ for 
$\lambda\searrow\lambda_c$. The trace of the energy-momentum
tensor $\theta_{\mu\nu}$ grows as $\frac{\theta_{\mu\mu}}{\Lambda^4}=\frac{\rho-3P}{\Lambda^4}=12\,\lambda$
for $\lambda\gg \lambda_c$. 
\end{theorem}
\begin{proof}
The Legendre transformation $\rho=T\frac{dP}{dT}-P$ between total energy
density $\rho$ and total pressure $P$, which follows from the
(ultraviolet regularized) partition
function formulated in terms of fundamental fields, needs to be honored
in the effective theory. Since there are implicit temperature
dependences in the parameters $|\phi|$ and $e$ of 
the effective theory for the coarse-grained fluctuations 
$\delta a_\mu$ the derivatives w.r.t. 
temperature of these parameters ought to cancel one another. A necessary
and sufficient condition for this to take place is $\partial_m
P=0$. Because one may neglect the quantum part, 
on the one-loop level $P$ and $\rho$ are determined by the thermal parts
of the propagators in Eqs.\,(\ref{prop12}) and (\ref{prop3}) (terms with
the factor $n_B$). They are given as follows:  
\begin{equation}
\label{Poneloop}
P(\lambda)=-\Lambda^4\left\{\frac{2\lambda^4}{(2\pi)^6}\left[2\bar{P}(0)+6\bar{P}(2a)\right]+2\lambda\right\}\,,
\end{equation}
\begin{equation}
\label{rhooneloop}
\rho(\lambda)=\Lambda^4\left\{\frac{2\lambda^4}{(2\pi)^6}\left[2\bar{\rho}(0)+6\bar{\rho}(2a)\right]+2\lambda\right\}\,,
\end{equation}
where $\bar{P}(y)\equiv\int_0^\infty
dx\,x^2\,\log\left[1-\exp(-\sqrt{x^2+y^2})\right]$ 
and $\bar{\rho}(y)\equiv\int_0^\infty
dx\,x^2\frac{\sqrt{x^2+y^2}}{\exp(\sqrt{x^2+y^2})-1}$. The evolution
equation (\ref{evoleqsu2}) follows by applying $\partial_{(aT)} P$ to the
expression in Eq.\,(\ref{Poneloop}) and setting the result equal to
zero. The right-hand side of the evolution equation (\ref{evoleqsu2})
indeed vanishes at $a=\infty$ (essential zero due to the exponential in
the integrand for the function $D(2a)$) and at $a=0$ (algebraic zero
since $D(0)$ exists: $D(0)=\frac{\pi^2}{6}$). Since the right-hand side
of Eq.\,(\ref{evoleqsu2}) is negative definite this equation is equivalent to
\begin{equation}
\label{evalambdasu2}
1=-\frac{24\lambda^3}{(2\pi)^6}\left(\lambda\frac{da}{d\lambda}+a\right)a\,D(2a)\,. 
\end{equation}
For $a\ll 1$ the Taylor expansion of the function $D(2a)$ can be
truncated at zeroth order\footnote{In \cite{DolanJackiw1974} it was
  shown that albeit the coefficients in the Taylor expansion of $D(y)$
  about $y=0$ diverge for orders larger or equal than quadratic this
  formal series can be resummed to a smooth function in $y$. For this process 
the zeroth-order coefficient serves as a boundary condition and thus is
relevant.}. This simplifies Eq.\,(\ref{evalambdasu2}) as  
\begin{equation}
\label{evalambdasu2sim}
1=-\frac{\lambda^3}{(2\pi)^4}\left(\lambda\frac{da}{d\lambda}+a\right)a\,, 
\end{equation} 
and the solution, subject to the initial condition $a(\lambda_i)=a_i\ll 1$ is
\begin{equation}
\label{solasu2} 
a(\lambda)=4\sqrt{2}\pi^2\lambda^{-3/2}\left(1-\frac{\lambda}{\lambda_i}\left[1-\frac{a_i\lambda_i^3}{32\pi^4}\right]\right)^{1/2}\,.
\end{equation}
Thus for $\lambda\ll\lambda_i$ the function $a(\lambda)$ runs into the
attractor $a(\lambda)=4\sqrt{2}\pi^2\lambda^{-3/2}$. Since $a\equiv\frac{m}{2T}=2\pi
e\lambda^{-3/2}$ there is a plateau $e\equiv\sqrt{8}\pi$ in this regime. Because the
attractor increases with
decreasing $\lambda$ the condition $a\ll 1$ will be violated at small 
temperatures. The estimate $14.61>\lambda_c$ is obtained by setting the 
attractor equal to unity. Since the true solution in this regime
will continue to grow with decreasing $\lambda$ (negative definiteness of
right-hand side of Eq.\,(\ref{evoleqsu2})) the right-hand side of 
Eq.\,(\ref{evoleqsu2}) will be exponentially suppressed. This 
verifies the behavior $a(\lambda)\propto -\log(\lambda-\lambda_c)$ for
$\lambda\searrow\lambda_c$ and
implies a logarithmic singularity at $\lambda_c$ also for
$e(\lambda)$. Numerically, one obtains 
$\lambda_c=13.87$. 

\noindent As for the (dimensionless) quantity $\frac{\theta_{\mu\mu}}{\Lambda^4}$ let
us consider the following function $h(\lambda)$
\cite{GiacosaHofmann2007}:
\begin{equation}
\label{hfunc}
h(\lambda)\equiv-\frac{\rho(\lambda)-3P(\lambda)}{4P^{\tiny\mbox{gs}}}\,.
\end{equation}
Expanding up to quadratic order in $a$, one has 
$h(\lambda)=1+\frac{\lambda^3 a^2(\lambda)}{4(2\pi)^4}$. Substituting
the attractor $a(\lambda)=4\sqrt{2}\pi^2\lambda^{-3/2}$, one 
arrives at $h(\lambda)=\frac32\,,\ (\lambda\gg\lambda_c)$, and the claim
follows by recalling that $P^{\tiny\mbox{gs}}=-4\pi\Lambda^3 T$.  
\end{proof}
The existence of an attractor to the evolution of $e$ with temperature
signals the decoupling of the high-temperature initial situation from
the low-temperature physics. According to Eq.\,(\ref{solasu2}) this
decoupling takes place in a power-like fashion in contrast to
perturbation theory. Also, the Yang-Mills scale $\Lambda$ is {\sl not}
determined by the initial value $a_i=2\pi e_i\lambda_i^{-3/2}$ (as it
would in perturbation theory), but it arises as a purely nonperturbative
integration constant owing to the nontrivial thermal ground state, see
Thm.\,\ref{lambdaemrg}. In a physics model the initial temperature
$T_i=\frac{\lambda_i\Lambda}{2\pi}$ is naturally given 
by the scale where the assumption of a smooth spacetime manifold
supporting the Yang-Mills theory breaks down. 
According to present consensus this scale is the Planck mass. The
constancy of $e$ for $a\ll 1$ signals that the magnetic charge $g=\frac{4\pi}{e}$ of a screened 
magnetic monopoles, liberated by a dissociating caloron/anticaloron of
large holonomy, is conserved during most of the evolution. 
For $\lambda\searrow\lambda_c$ both the magnetic charge $g$ and the mass
$M_{\tiny\mbox{mon}}\sim\frac{4\pi^2 T}{e}$ of a screened magnetic
monopole vanish, see Rem.\,\ref{temperatureLLKvB}. 

In a completely analogous way one obtains for SU(3) the evolution
equation 
\begin{equation}
\label{evolsu3dec}
\partial_a\lambda=-\frac{12\lambda^4
  a}{(2\pi)^6}\frac{D(a)+2\,D(2a)}{1+\frac{12\lambda^3a^2}{(2\pi)^6}(D(a)+2\,D(2a))}\,.
\end{equation}
The attractor for $a\ll 1$ reads \cite{GiacosaHofmann2007}
$a(\lambda)=\frac{4}{\sqrt{3}}\pi^2\lambda^{-3/2}$, and the plateau
value is $e\equiv \frac{4}{\sqrt{3}}\pi$. The numerical value for
$\lambda_c$ is $\lambda_c=9.475$, and one obtains
$\theta_{\mu\mu}=24\pi\Lambda^3\,T=12\,\lambda\Lambda^4$ for $\lambda\gg\lambda_c$. 

\subsection{Radiative corrections}

Here we discuss the systematic computation of radiative corrections in terms 
of a respective loop expansion within the effective theory for the deconfining
phase. This loop expansion has little resemblence with 
its perturbative counterpart. Each nonvanishing diagram is infrared and ultraviolet 
finite. While the former property is due to the nonpertburbative
emergence of (quasiparticle-)mass on tree level (adjoint Higgs
mechanism) the ultraviolet finiteness follows from the existence of a
scale $|\phi|$ of maximal resolution: In a physical gauge quantum 
fluctuations are constrained to maximal hardness $|\phi|^2$ by the 
thermal ground state, and their action is small as compared 
to that of thermal (on-shell) modes. Frankly speaking, this is the
reason for the rapid converence of loop expansions. 
We also exhibit two calculational examples:
The on-shell one-loop polarization tensor of the massless mode and the
dominant two-loop correction to the pressure.  

After a euclidean rotation $p_0\to ip_4\,,(p_0\ \mbox{and}\ p_4\
\mbox{real})$ the second condition in (\ref{cc1}) reads 
$|p^2+4e^2|\phi|^2|\le|\phi|^2$. Since this is never true in the
attractor regime because $e\ge\sqrt{8}\pi$ for SU(2) and
$e\equiv\frac{4}{\sqrt{3}}\pi$ for SU(3) one concludes that massive 
quasiparticles do propagate thermally only. Alternatively, staying in 
Minkowskian signature, the four-momentum squared of a massive mode needs 
tuning to the mass squared of about one part in three-hundred for SU(2)
and similarily for SU(3). Again, this implies 
that a massive quasiparticle propagates over long 
distances and thus thermalizes. Also massive 
quasiparticles cannot be created by quantum processes because this would
invoke momentum tranfers of at least twice their mass. But this is about
35 times larger than the maximally allowed resolution in 
the effective theory.\vspace{0.2cm}\\ 
{\sl An irreducible bubble diagram is defined by the property that no
one-particle reducible diagram for the polarization tensor is created by cutting in any possible way
an internal line of the diagram.}\vspace{0.2cm}\\ 
It was argued in \cite{Hofmann2006} that, when computing bubble diagrams contributing to
the thermal pressure, the resummation of polarization tensors, subject
to an insertion of an irreducible bubble diagram, avoids the occurrence 
of pinch singularities (powers of delta functions) due to a relaxation
to finite-width spectral functions in the thermal parts of the
so-obtained real-time propagators. Technically, 
the procedure for resumming polarization tensors is to 
compute them in real time subject to the constraints (\ref{cc1}) and (\ref{cc2}), perform an analytic continuation to imaginary time in the
external momentum, carry out the resummation, and finally continue back
to real time.\vspace{0.2cm}\\ 
{\sl For a four momentum $p=(p_0,\vec{p})$ circulating in a loop we refer to  
$p_0$ and $|\vec{p}|$ as the two independent radial loop momenta. We
denote by $\tilde{K}$ the number of independent radial loop momenta in a
given irreducible bubble diagram.}
\vspace{0.2cm}\\ 
{\sl Independent hypersurfaces $H_i\,,\ (i=1,\cdots,h\le\tilde{K})\,,$ 
in the $\tilde{K}$-dimensional space ${\mathbbm R}^{\tilde{K}}$ are
defined by the property that in a whole environment $U$ of their
intersection $\bigcap_{i=1}^h H_i$ the normal vectors normal vectors $\hat{n}_i$ to $H_i$ 
(computed anywhere on $U\cap H_i$) are linearly independent.}
\begin{remark}
\label{h=K}
If $h=\tilde{K}$ then it follows that $\bigcap_{i=1}^{\tilde{K}} H_i$ is
a set of discrete points.
\end{remark}
\begin{proposition}
\label{K4}
For an irreducible bubble diagram containing only $V_4$-many
four-vertices and no three-vertices the
number $K$ of independent constraints on the loop momenta is estimated
as $K\ge\frac{7}{2}\,V_4$. 
\end{proposition}
\begin{proof}
The number $I$ of internal lines in such a diagram is $I=2\,V_4$
\cite{Weinberg}. Because of (\ref{cc1}) there are thus $2\,V_4$ 
constraints on propagating momenta and according to (\ref{cc2}) at least 
$\frac32\,V_4$ constraints on momentum transfers in vertices. (For this
estimate pair the four-vertices in the diagram.) Together this gives
$K\ge\frac{7}{2}\,V_4$. 
\end{proof}
\begin{proposition}
\label{K3}
For an irreducible bubble diagram containing only $V_3$-many
three-vertices and no four-vertices the
number $K$ of independent constraints on the loop momenta is given as $K=\frac{3}{2}\,V_3$. 
\end{proposition}
\begin{proof}
The number $I$ of internal lines in such a diagram is $I=\frac32\,V_3$
\cite{Weinberg}. Because of (\ref{cc1}) there are thus $\frac32\,V_3$ 
constraints on propagating momenta. No additional constraints arise 
because the momentum transfer in the vertex, induced by two external
legs, coincides by momentum conservation with the momentum of the third
leg. Thus $K=\frac32\,V_3$. 
\end{proof}
\begin{proposition}
\label{Ktilde4}
For an irreducible bubble diagram containing only $V_4$-many
four-vertices and no three-vertices the
number $\tilde{K}$ of independent radial loop momenta is given as
$\tilde{K}=2\,V_4+2$.
\end{proposition}
\begin{proof}
The Euler characteristic for a spherical polyhedron reads $2=V-I+F$
where $V$ is the number of vertices, $I$ the number of edges, and $F$
the number of faces. Since a connected bubble diagram {\sl is} a spherical 
polyhedron with one face removed the identification 
of $F=L+1$, where $L$ denotes the number
of loops (or left-over faces), yields $L=I-V+1$. Combining this with
$I=2\,V_4$ and using the fact that $\tilde{K}$ is twice the number $L$
proves the claim.   
\end{proof}
\begin{proposition}
\label{Ktilde3}
For an irreducible bubble diagram containing only $V_3$-many
three-vertices and no four-vertices the
number $\tilde{K}$ of independent radial loop momenta is given as
$\tilde{K}=V_3+2$.
\end{proposition}
\begin{proof}
Combining $L=I-V+1$ with
$I=\frac32\,V_3$ and using the fact that $\tilde{K}$ is twice the number
$L$ proves the claim. 
\end{proof}
\begin{corollary}
\label{ratiosKtideK}
From Props.\,\ref{K4},\ref{K3},\ref{Ktilde4}, and \ref{Ktilde3} one
concludes that
\begin{equation}
\label{K4erat}
\frac{\tilde{K}}{K}\le\frac47\left(1+\frac{1}{V_4}\right)\,,\ \
(\mbox{four-vertices only})\,;\ \ \ 
\frac{\tilde{K}}{K}=\frac23\left(1+\frac{2}{V_3}\right)\,,\ \
(\mbox{three-vertices only})\,. 
\end{equation}
Since any subdiagram (obtained by cutting more than one internal
line) of an irreducible bubble diagram is again 
irreducible it follows that at a fixed number $V$ of vertices with
$V=V_4+V_3\ge 2$ the ratio $\frac{\tilde{K}}{K}$ 
is minimal for $V_3=0$ and maximal for
$V_4=0$.
\end{corollary}
If the constraints (\ref{cc1}) and (\ref{cc2}) were {\sl equations} 
and not inequalities one would conclude from Cor.\,\ref{ratiosKtideK} 
that the intersection of the independent hypersurfaces $H_i$ specified by 
them would be empty for a number $L$ of loops greater than a finite
number $L_{\tiny\mbox{max}}$. From
$\frac{\tilde{K}}{K}\le\frac23\left(1+\frac{2}{V_3}\right)=1$ we have
$V_{\tiny\mbox{max}}=4$ which by virtue of the proof to
Prop.\,\ref{Ktilde4} implies that $L_{\tiny\mbox{max}}\le 5$. 

For completeness let us investigate the generalization of the Euler characteristic for a 
spherical polyhedron without any handles to the situation of nonplanar 
bubble diagrams. Notice that the latter can be considered spherical polyhedra with one face removed and a nonvanishing number 
of handles added (genus $g>0$):
\eqb
\label{gen}
V-I+L+1=2\ \ \longrightarrow \ \ V-I+L+1=2-2g\,,
\eqe
where again $I$ is the number of internal lines, $L$ the number of loops, and $g$ represents the genus of 
the polyhedral surface (the number of handles). Notice that the right-hand side of the right-hand 
side equation is the full Euler-L' Huilliers characteristics. Reasoning as above but now based on the 
general situation of $g\ge 0$ expressed by the right-hand side equation in Eqs.\,(\ref{gen}), we arrive at
\eab
\label{ratiosgen}
\frac{\tilde{K}}{K}&\le&\frac47\left(1+\frac{1}{V_4}\left(1-2g\right)\right)\,,\ \ \ (V=V_4)\,,\nonumber\\ 
\frac{\tilde{K}}{K}&\le&\frac23\left(1+\frac{2}{V_3}\left(1-2g\right)\right)\,,\ \ \ (V=V_3)\,.
\eae
According to Eqs.\,(\ref{ratiosgen}) the demand $\frac{\tilde{K}}{K}\le 1$ for a compact support of the 
loop integrations is always satisfied for $g\ge 1$ since the number of vertices needs to be positive: 
$V_4\ge 0$ and $V_3\ge 0$. Recall that at $g=0$ this is true 
only for $V_4\ge 2$ and $V_3\ge 6$, respectively. We thus conclude that bubble diagrams of a topology deviating from 
planarity are much more severely constrained than their planar counterparts. 

Because other thermodynamic 
quantities are related to the pressure by (successive) Legendre 
transformations this would imply their exact calculability as well. 
But since (\ref{cc1}) and
(\ref{cc2}) are inequalities the associated hypersurfaces $H_i$ are fattened, and the
situation is less clear cut. However, for large 
$V_3$ the ratio $\frac{\tilde{K}}{K}$ approaches the value $\frac23$,
which is considerably smaller than unity, in
a powerlike-way suggesting that $\bigcap_{i=1}^K H_i=\emptyset$ for
sufficiently large but finite $K$ (or $V_3$ or $L$). 
Here $H_i$ now refers to a fattened hypersurface. We thus arrive at the
following conjecture:\vspace{0.2cm}\\   
\noindent{\sl The loop expansion of the pressure in the effective theory for the 
deconfining phase terminates at a finite loop order.}
\begin{remark}
\label{su3le}
The argument presented in favor of the truth of this conjecture apply to both the
SU(2) and the SU(3) case. In \cite{KavianiHofmann2007} we have observed
that for the SU(2) case the three-loop
irreducible bubble diagram with $V_4=2$ and $V_3=0$, containing two
internal lines with massless and two internal lines with massive
particles, vanishes identically.\end{remark} 
\begin{example}
\label{polarizationtensor} 
Here we provide a one-loop example for a typical radiative 
correction: the polarization tensor $\Pi_{\mu\nu}$ for the massless 
mode with $p^2=0$ in the effective theory for deconfining SU(2) Yang-Mills 
thermodynamics \cite{SHG2006-1}. Without restriction of generality one may
assume that $\vec{p}$ points into the 3-direction. Then 
the only nontrivial entries in $\Pi_{\mu\nu}$ are
$\Pi_{11}=\Pi_{22}\equiv G(p_0=|\vec{p}|,\vec{p},T,\Lambda)$. One easily checks 
that only the tadpole diagram with the massive modes ($a=1,2$)
circulating in the loop contributes for one-shell external momentum
($p^2=0$). That is, for $p^2=0$ and on the one-loop level there is no imaginary part in the
screening function\footnote{On the two-loop level and at $p^2=0$ $G$ receives an
  imaginary contribution whose modulus, however, is strongly suppressed
  as compared to the modulus of the one-loop result, for the ratio 
  of two-loop to one-loop contributions to the pressure see \cite{SHG2006-1}.}
$G$. By a (lengthy) routine computation, which takes into account the
constraints\footnote{The three constraints in (\ref{cc2}) collapse onto
  one constraint for this diagram.} (\ref{cc2}) and recalls that massive modes
propagate thermally only, one obtains the following result:
\begin{figure}
\begin{center}
\leavevmode
\leavevmode
\vspace{6.1cm}
\includegraphics{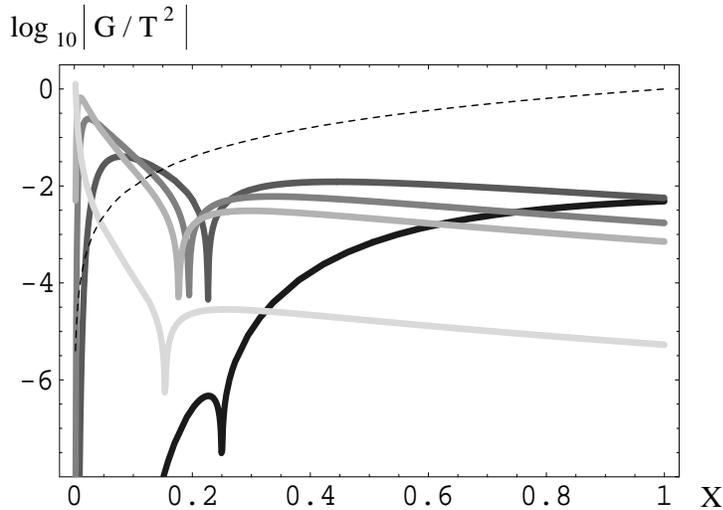}
\end{center}
\caption{\protect{\label{Fig-2}}$\left|\frac{G}{T^2}\right|$ as a function of $X\equiv\frac{|\vec{p}|}{T}$ 
for $\lambda=1.12\,\lambda_{c}$ (black), $\lambda=2\,\lambda_{c}$ (dark grey), 
$\lambda=3\,\lambda_{c}$ (grey), $\lambda=4\,\lambda_{c}$ (light grey), $\lambda=20\,\lambda_{c}$ 
(very light grey). The dashed curve is a 
plot of the function $f(X)=2\log_{10}X$. 
There is screening to the left ($G>0$) and antiscreening ($G<0$) to the right of the
cusps. The massless mode is strongly screened at 
$X$-values for which $\log_{10}\left|\frac{G}{T^2}\right|>f(X)$ ($\frac{\sqrt{G}}{T}>X$), 
that is, to the left of the dashed line. For $\lambda=\lambda_c$ the
function $G$ vanishes identically.}      
\end{figure}
\begin{equation}
\label{Goflambda1}
\frac{G}{T^2}=\left[\int_{-\infty}^{\xi_m(X,\lambda)}d\xi\,
\int_{\rho_m(X,\xi,\lambda)}^{\rho_M(X,\xi,\lambda)} d\rho+
\int_{\xi_m(X,\lambda)}^{\xi_M(X,\lambda)}
d\xi\,\int_0^{\rho_M(X,\xi,\lambda)} d\rho\right]\times\nonumber
\end{equation}
\begin{equation}
\label{Goflambda2}
\hspace{1.3cm}e^2(\lambda)\lambda^{-3}\left(-4+\frac{\rho^2}{4e^2(\lambda)}\right)\,\rho\,\frac{n_B\left(2\pi \lambda^{-3/2}\sqrt{\rho^2+\xi^2+4e^2(\lambda)}\right)}
{\sqrt{\rho^2+\xi^2+4e^2(\lambda)}}\,,
\end{equation}
where the dimensionless quantities $\xi_m,\xi_M,\rho_m,$ and $\rho_M$ are
given as 
\begin{equation}
\label{cutoffspol1}
\xi_{\stackrel{m}M}(X,\lambda)\equiv \frac{\pi}{2X}\,\frac{4e^2\mp
  1}{\lambda^{3/2}}-2\,
\frac{X}{\pi}\,\lambda^{3/2}\,
\frac{e^2}{4e^2\mp 1}\,, 
\end{equation}
\begin{equation}
\label{cutoffspol2}
\rho_{\stackrel{m}M}(X,\xi,\lambda)\equiv\sqrt{\left(\frac{\pi}{X}\right)^2\,\frac{(4e^2\mp
    1)^2}{\lambda^3}-
\frac{2\pi}{X}\,\frac{4e^2\mp 1}{\lambda^{3/2}}\xi-4e^2}\,,
\end{equation}
the dimensionless quantity $X$ is defined as
$X\equiv\frac{|\vec{p}|}{T}$, and $e(\lambda)$ follows from the solution
$a(\lambda)$ of the evolution equation (\ref{evalambdasu2}) 
by virtue of $e(\lambda)=\frac{a(\lambda)}{2\pi}\,\lambda^{3/2}$. A plot of $\log_{10}\frac{G}{T^2}$ in
dependence of $X$ for various values of $\lambda$ is presented in
Fig.\,\ref{Fig-2}. By virtue of the constraints expressed in (\ref{cutoffspol1}) and
(\ref{cutoffspol2}) it is straight-forward to show that the function 
$G$ possesses an essential zero at $X=0$.  
\end{example}
\begin{example}
\label{2loopdom} 
In Fig.\,\ref{Fig-3} the dependence on temperature of the ratio of the 
dominant two-loop diagram for the pressure (with $V_3=2$ and $V_4=0$)
and the ground-state subtracted one-loop result is depicted. We refrain 
from quoting formulas and refer the reader to \cite{SHG2006-1}. 
\begin{figure}
\begin{center}
\leavevmode
\leavevmode
\vspace{5.9cm}
\includegraphics{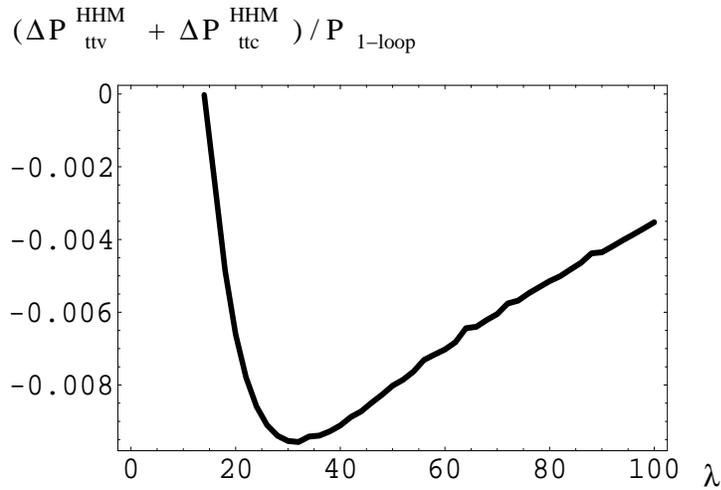}
\end{center}
\caption{\protect{\label{Fig-3} The ratio of the dominant two-loop 
correction to the pressure and the ground-state subtracted one-loop 
result as a function of $\lambda$.}}      
\end{figure}
\end{example}
\begin{remark}
\label{hierachydia}
Combining the results of \cite{Hofmann2005L}, \cite{SHG2006-1}, and
\cite{KavianiHofmann2007}, the ratio of two-loop corrections to the
pressure to the ground-state subtracted one-loop result is, depending on temperature, at most
$\sim 10^{-2}$ and the ratio of three-loop corrections to the
ground-state subtracted one-loop
result at most $\sim 2\times 10^{-7}$.  
\end{remark}

\section{Preconfining phase\label{PP}} 

\subsection{Thermal ground state: Interacting magnetic monopoles/antimonopoles}

The preconfining phase is the mediator between deconfinement at high
temperatures and the low-temperature confining phase where (dual) gauge fields do not
propagate. This phase occupies a very narrow region in the phase diagrams
of SU(2) and SU(3) Yang-Mills thermodynamics. This and the fact that
lattice simulations operate at a finite spatial volume and thus have
difficulty to exhaustively capture the important long-range correlations
inherent to the monopole-antimonopole condensing ground state are the 
reasons why the preconfining phase has escaped its detection in 
numerical experiments \cite{TeperI,TeperII,TeperIII}. 
We know that at $\lambda_c$ one direction in the
SU(2) algebra remains precisely massless and thus 
propagates. At the same time we know that screened magnetic monopoles
and antimonopoles become massless at $\lambda_c$ and thus are prone 
to condensation. Even though the monopole-antimonopole condensate 
induces a dynamical breaking of the (dual) U(1) gauge 
symmetry this does not lead to the immediate decoupling of 
the dual gauge mode. Why and at what point this happens 
and how the deconfining ground state relates to the
monopole-antimonopole condensate (tunneling) when lowering 
the temperature is subject to insightful analysis. On average, 
the process of gradually generating an extra 
polarization of the dual gauge mode by tunneling transitions 
between the deconfining ground state, characterized by 
a short-lived and small caloron or anticaloron holonomy, 
and the monopole-antimonopole condensate,
characterized by a stable and large caloron or anticaloron holonomy 
\cite{Diakonov2004,Muller-Preussker2006,DiakonovPetrov2007} 
appears to be relevant when addressing the emergence of intergalactic
magnetic fields in applying SU(2) Yang-Mills thermodynamics to
describe thermalized photon propagation, 
see \cite{GH2005,SHG2006-1,SHG2006-2,SHGS2007,Hofmann2005L,Hofmann2005S}.

In this section a derivation of the thermal ground state in the
preconfining phase is given where magnetic monopoles and their antimonopoles 
are pairwise condensed. The process of condensation is extremely subtle 
microscopically: As temperature approches $\lambda_c$ from above, 
the screening of a given preexisting 
magnetic monopole and its antimonopole (liberated by the dissociation of a
large-holonomy caloron or anticaloron) is suddenly enhanced (logarithmic
pole in $e(\lambda)$). Although this rapidly suppresses their 
mass and magnetic charge it does not yet lead to the formation of a
stable condensate. For $\lambda<\lambda_c$ the average
caloron-anticaloron 
holonomy gradually increases 
with decreasing temperature (supercooling) giving rise 
to more frequent caloron and anticaloron dissociation processes. As a
consequence, the magnetic-charge screening of a 
given monopole increasingly is due to alike monopoles and antimonopoles
(and decreasingly due to annihilating monopoles and antimonopoles inside
a small-holonomy caloron/anticaloron) in an ever more stable condensate. 
Although the derivation of a complex scalar field describing the
monopole-antimonopole condensate only relies on the 
limit of total screening $e\to\infty$, which takes place at $\lambda_c$,
the formation of a stable condensate is seen to occur at a slightly
smaller temperature.  

\begin{remark}
\label{scaleinvariance}
The condensate of monopoles and antimonopoles starts to form at 
$\lambda_c$, where $e=\infty$, and the Yang-Mills system `forgets' about the
existence of the mass scale $\Lambda$ since the mass of monopoles and
antimonopoles vanishes by total screening and since there is no
interaction between them because the magnetic 
coupling $g=\frac{4\pi}{e}$ vanishes. Also, since screened monopoles and 
antimonopoles are at rest w.r.t. the heat bath (previously 
created by dissociating large-holonomy calorons and anticalorons 
\cite{Diakonov2004}) no kinetic energy of their motion exists. 
Thus, if, after an appropriate spatial coarse-graining, 
the condensate is described by a spatially homogeneous 
field $\varphi$, then this field by itself must be BPS saturated, that is, its
energy-momentum tensor (or, after integrating out any dependence 
on space, its euclidean energy density) vanishes identically.
\end{remark}
{\sl The potential for the formation of a monopole-antimonopole condensate
opens up for $e\to\infty$. After an appropriate spatial 
coarse-graining and on the level of no interactions between monopoles
and/or antimonopoles this condensate is, in the euclidean formulation, described 
by an inert, spatially homogeneous, and 
BPS saturated complex scalar field 
$\varphi=|\varphi|\exp\left[\pm 2\pi i\frac{\tau}{\beta}\right]$ where
$|\varphi|=\sqrt{\frac{\bar{\Lambda}^3\beta}{2\pi}}$, and $\bar{\Lambda}$ is
an arbitrary mass scale. Taking
interactions between monopoles and antimonopoles into 
account, the thermal ground state is described by $\varphi$ and the
pure-gauge configuration $a^{D,\tiny\mbox{gs}}_\mu=\mp\delta_{\mu 4}\frac{2\pi}{g\beta}$ of the dual abelian 
gauge field $a^D_\mu$. Here $g$ is the magnetic coupling, and the
ground-state energy density $\rho^{\tiny\mbox{gs}}$ and pressure
$P^{\tiny\mbox{gs}}$ are given as
$\rho^{\tiny\mbox{gs}}=-P^{\tiny\mbox{gs}}=\pi\bar{\Lambda}^3T$.}\vspace{0.2cm}\\ 
$\mbox{}$\\ 
{\sl complex scalar field $\varphi$}:\\ 
For the effective theory (\ref{fullactden}) in unitary gauge 
out of the three directions $a^{1,2,3}_\mu$ in the Lie algebra 
only $a^{3}_\mu$ propagates for $e\to\infty$, the other 
two gauge modes decouple because their quasiparticle 
mass diverges. For $\lambda\le\lambda_c$ we 
define the dual gauge field $a^D_\mu$ by the coarse-grained version of
$a_\mu^{3}$. Since $a_\mu^{3}$ is a free field for 
$\lambda\le\lambda_c$ this coarse-graining is 
trivial. In particular no (local or nonlocal) composites of the field 
$a^D_\mu$ may propagate in the effective theory for 
the preconfining phase, and $a_{\mu}^D\to
a_{\mu}^D+\frac{i}{g}\Omega\partial_\mu\Omega^{\dagger}$ under a gauge
transformation with $\Omega\in\,$U(1). Because of the rotational symmetry of the thermal 
system the monopole-antimonopole condensate is 
described by a scalar field $\varphi$. Since $\varphi$ by 
itself is BPS saturated it would be irrelevant to the thermodynamics of
the preconfining phase if it was a U(1) gauge 
singlet (real scalar). But then the 
only option for coupling $\varphi$ to $a_{\mu}^D$ in a 
gauge-invariant way is the transformation law
$\varphi\to\Omega^\dagger\varphi$ (complex scalar field).\\ 
{\sl $\varphi$'s phase}:\\  
The (dimensionless) phase $\theta$ with $\varphi=|\varphi|\exp[i\theta]$
is defined by the geometrically and thermally averaged 
magnetic flux $\bar{F}_{\pm,\tiny\mbox{th}}$ through a two-dimensional sphere, $S_{2,R=\infty}$, 
of vanishing curvature (infinite radius $R$) induced by a
monopole-antimonopole system\footnote{A monopole is {\sl correlated} with its 
antimonopole in the sense that the former owes its existence to the
latter and vice versa since their origin is the 
charge separation enabled by the strong deformation of 
a small-holonomy caloron/anticaloron.} at 
zero-momentum and $e\to\infty$. In accord with
Rem.\,\ref{scaleinvariance} this is the only possible 
definition of a dimensionless quantity which 
does not make any reference to a scale. Consider a system of
a zero-momentum monopole and its zero-momentum antimonopole. In unitary
gauge, where independently of position 
the adjoint Higgs field of the monopole or antimonopole configuration 
points into a fixed direction in the SU(2) 
algebra and Dirac strings compensate for the magnetic flux through a
closed surface surrounding the monopole or antimonopole, we introduce 
unit vectors $\hat{x}_m$ and $\hat{x}_a$ for monopole and antimonopole,
respectively. These vectors signal the direction of the Dirac strings (both
pointing away from the respective center of charge). Let $\delta\equiv
\angle(\hat{x}_m,\hat{x}_a)$ and both monopole and 
antimonopole be placed on the same side of $S_{2,R=\infty}$. (It can
easily be checked below that this is no restriction of generality.) 
Now a single monopole or a single antimonopole, whose Dirac string does
not pierce $S_{2,R=\infty}$, would induce a magnetic flux $F_{\pm}$ through
$S_{2,R=\infty}$ of $F_{\pm}=\pm\frac{4\pi}{e}$. It is then easy to see \cite{Hofmann2005L} that, 
(geometrically) averaging over all directions of $\hat{x}_m$ and $\hat{x}_a$ at a given angle 
$\delta$, the flux $\bar{F}_{\pm}(\delta)$ of the monopole-antimonopole system is given as 
\begin{equation}
\label{geoflux}
\bar{F}_{\pm}(\delta)=\pm\frac{\delta}{2\pi}\frac{4\pi}{e}=\pm\frac{2\delta}{e}\,,\
\ \ \ \ (0\le\delta\le\pi)\,.        
\end{equation}
After screening the mass $M_{m+a}$ 
of the monopole-antimonopole system is given
as \cite{LeeLu1998} $M_{m+a}=\frac{8\pi^2}{e\beta}$. Thus, coupling this system to the heat
bath, the thermally averaged flux $\bar{F}_{\pm,\tiny\mbox{th}}(\delta)$
reads
\begin{equation}
\label{thermgeoflux}
\bar{F}_{\pm,\tiny\mbox{th}}(\delta)=4\pi\int
d^3p\,\delta^{(3)}(\vec{p})\,n_B(\beta E(\vec{p}))\,
\bar{F}_{\pm}(\delta)\,,
\end{equation}
where $E(\vec{p})\equiv\sqrt{M_{m+a}^2+\vec{p}^2}$, and
  $n_B(x)\equiv\frac{1}{\exp[x]-1}$ denotes the Bose function. Since
\begin{equation}
\label{expdenBose}
\lim_{\vec{p}\to
  0}\left(\exp\left[\beta\sqrt{M^2_{m+a}+\vec{p}^2}\right]-1\right)=\frac{8\pi^2}{e}\left(1+
\frac12\frac{8\pi^2}{e}+\frac16\left(\frac{8\pi^2}{e}\right)^2+\cdots\right)\,
\end{equation}
one finally has
\begin{equation}
\label{finflux}
\lim_{e\to\infty}\bar{F}_{\pm,\tiny\mbox{th}}(\delta)=\pm\frac{\delta}{\pi}\equiv
\frac{\theta}{2\pi}\,,\ \ \ \ (0\le\delta\le\pi)\,.
\end{equation}
Since the angle $\delta$, which, after eliminating any 
dependence on space by the inclusion of vanishing 
spatial momentum into the average thermal flux, see
Eq.\,(\ref{thermgeoflux}), has lost its original geometric 
meaning but still ought to parametrize a periodic
situation, we may set
$\pm\frac{\delta}{\pi}=\pm\frac{\tau}{\beta}\,,\ \
(0\le\tau\le\beta)$. That is, periodicity of the 
flux in dependence of an angle in the monopole-antimonopole condensate
is, after spatial coarse-graining, 
promoted to the periodicity in euclidean time of the 
associated, spatially homogeneous, BPS saturated field $\varphi$. Therefore, we have 
\begin{equation}
\label{taudepvarphi}
\varphi=|\varphi|\exp\left[\pm 2\pi i\frac{\tau}{\beta}\right]\,.
\end{equation}
Since $|\varphi|$ is spatially homogeneous with the same justification
as for the field $\phi$ in the deconfining phase the field
$\varphi$ is annihilated by the linear differential
operator\footnote{By a global U(1) gauge rotation a phase shift
  $\tau\to\tau+\tau_0$ can be introduced (global rotation of the Dirac
  strings), 
and $|\varphi|$ so far is an 
undetermined normalization. This freedom spans a two-dimensional vector 
space which coincides with the kernel $\bar{\cal K}$ of $\bar{\cal
  D}$ and thus determines $\bar{\cal
  D}$ uniquely.} $\bar{\cal
  D}\equiv\partial_\tau^2+\left(\frac{2\pi}{\beta}\right)^2$:
\begin{equation}
\label{Dannihilates}
\bar{\cal
  D}\varphi=\bar{\cal
  D}\varphi^*=0\,,
\end{equation}
where $\varphi^*$ denotes the complex conjugate of $\varphi$.\\  
{\sl $\varphi$'s modulus}:\\ 
No explicit temperature dependence may appear in the eucliden action for the
field $\varphi$ on the level of noninteracting monopoles and 
antimonopoles and $e\to\infty$. According to Eq.\,(\ref{Dannihilates})
and because of gauge invariance one may thus write
\begin{equation}
\label{varphipotential}
S_\varphi=\int_0^\beta d\tau\int
d^3x\left(\frac12\partial_\tau\varphi^*\partial_\tau\varphi+\frac12 V(|\varphi|^2)\right)\,, 
\end{equation}
where $V(|\varphi|^2)$ is a to-be-determined gauge-invariant potential
and $|\varphi|^2=\varphi^*\varphi$. By virtue of 
Eq.\,(\ref{taudepvarphi}) the Euler-Lagrange equation, which follows 
from the action (\ref{varphipotential}), reads
\begin{equation}
\label{ELvarphi}
\partial^2_\tau\varphi=\frac{\partial V(|\varphi|^2}{\partial
  |\varphi|^2}\varphi\ \ \ \ \ 
\stackrel{\tiny\mbox{Eq.\,(\ref{taudepvarphi})},\varphi\not=0}
\Longleftrightarrow\ \ \ \ \ \left(\frac{2\pi}{\beta}\right)^2=-\frac{\partial V(|\varphi|^2)}{\partial
  |\varphi|^2}\,. 
\end{equation}
On the other hand, the field $\varphi$ is BPS
saturated (vanishing of the euclidean
energy density). Eq.\,(\ref{varphipotential}) and Eq.\,(\ref{taudepvarphi}) thus implies that
\begin{equation}
\label{BPSvarphi}
|\varphi|^2\left(\frac{2\pi}{\beta}\right)^2-V(|\varphi|^2)=0\,.
\end{equation}
Together, Eqs.\,(\ref{ELvarphi}) and (\ref{BPSvarphi}) yield
\begin{equation}
\label{eompotvarphi}
\frac{\partial V(|\varphi|^2)}{\partial|\varphi|^2}=-\frac{V(|\varphi|^2)}{|\varphi|^2}\,.
\end{equation}
The solution to the first-order equation (\ref{eompotvarphi}) reads
\begin{equation}
\label{solpotvarphi}
V(|\varphi|^2)=\frac{\bar{\Lambda}^6}{|\varphi|^2}\,,
\end{equation}
where $\bar{\Lambda}$ is a mass scale which appears as a constant of 
integration. Substituting Eq.\,(\ref{solpotvarphi}) into
Eq.\,(\ref{BPSvarphi}) yields
\begin{equation}
\label{varphimod}
|\varphi|=\sqrt{\frac{\bar{\Lambda}^3}{2\pi T}}=\sqrt{\frac{\bar{\Lambda}^3\beta}{2\pi}}\,.
\end{equation} 
The quantity $|\varphi|$ sets the scale of maximal resolution in the
effective theory. An $S_{2,R=|\varphi|^{-1}}$ separating a monopole in
the interior from its antimonopole in the exterior (or vice versa)
experiences the same magnetic flux as an $S_{2,R=\infty}$ since in
the condensate the monopole-antimonopole distance and their core-size 
is nil. Thus monopole and antimonopole cannot probe the finite curvature
of $S_{2,R=|\varphi|^{-1}}$ and the infinite-surface limit is trivially
saturated in the spatial coarse-graining.\\  
{\sl $\varphi$'s inertness}:\\ 
By virtue of Eqs.\,(\ref{solpotvarphi}) and (\ref{varphimod}) one has 
\begin{equation}
\label{varphimodinert}
\partial^2_{|\varphi|} V(|\varphi|^2)=6\,\bar{\lambda}^3|\varphi|^2=24\,\pi^2\,T^2\,,
\end{equation}
where $\bar{\lambda}\equiv\frac{2\pi T}{\bar{\Lambda}}$. We will show 
below that $\bar{\lambda}\ge 7.075$. Thus the field $\varphi$ neither 
fluctuates quantum mechanically nor thermally.\\ 
{\sl full action and $a^{D,\tiny\mbox{gs}}_\mu$:}\\  
Since the field $a_\mu^3$ does not interact with itself the 
coarse-grained field $a_\mu^D$ obeys the same form of the 
action. Also, local U(1) gauge invariance dictates that $\partial_\tau\to
{\cal D}_\mu\equiv\partial_\mu+ig a_{\mu}^D$. The effective action for the
preconfining phase thus reads
\begin{equation}
\label{Spreconf}
S=\int_0^\beta d\tau\int d^3x\,\left[\frac14 G^D_{\mu\nu}
  G^D_{\mu\nu}+\frac12 ({\cal D}_\mu\varphi)^*{\cal
    D}_\mu\varphi+
\frac12\frac{\bar{\Lambda}^6}{|\varphi|^2}\right]\,,
\end{equation}
where $G^D_{\mu\nu}\equiv\partial_\mu a^D_\nu-\partial_\nu
a^D_\mu$. Making use of the inertness of the field $\varphi$, the
Euler-Lagrange equations, which follow from the action (\ref{Spreconf}),
are given as 
\begin{equation}
\label{ELdualgf}
\partial_{\mu} G^D_{\mu\nu}=ig\left[({\cal D}_\nu\varphi)^*\varphi-\bar{\varphi}{\cal D}_\nu\varphi^*\right]\,.
\end{equation}
By virtue of ${\cal D}_\nu\varphi=0$ the pure-gauge configuration
$a_\mu^{D,\tiny\mbox{gs}}=\mp\frac{2\pi}{g\beta} \delta_{\mu 4}$
solves Eq.\,(\ref{ELdualgf}). Inserting $a_\mu^{D,\tiny\mbox{gs}}$ and
$\varphi$ into (\ref{Spreconf}) one reads off the ground-state energy
density and pressure as
$\rho^{\tiny\mbox{g.s.}}=-P^{\tiny\mbox{gs}}=\pi\bar{\Lambda}^3T$.\vspace{0.2cm}\\  
{\sl The Polyakov loop, evaluated on the dual gauge-field configuration 
$a_\mu^{D,\tiny\mbox{gs}}$, is unity independently of the choice of
admissible gauge.}\vspace{0.2cm}\\     
Let us show this. The field $\varphi$ remains periodic 
under $\varphi\to\Omega^\dagger\varphi$ (admissible change of gauge) 
if and only if $\Omega=\exp\left[i\left(2\pi
  n\frac{\tau}{\beta}+\alpha(\vec{x})\right)\right]$ where
$n\in\mathbbm Z$, and $\alpha$ is a real
function of space only. Hence 
$a^D_\mu\to a^D_\mu+\frac{2\pi
  n}{g\beta}\delta_{\mu 4}+\frac{\partial_j\alpha(\vec{x})}{g}\delta_{\mu j}$ 
under $\Omega$, and thus the
periodicity of $a^D_\mu$ is (trivially) assured. (Here $j=1,2,3$.) 
In particular,
\begin{equation}
\label{periodadual} 
a_\mu^{D,\tiny\mbox{gs}}\to(\mp\frac{2\pi}{g\beta}+\frac{2\pi n}{g\beta})\delta_{\mu
  4}+\frac{\partial_j\alpha(\vec{x})}{g}\delta_{\mu j}=
\frac{2\pi (n\mp 1)}{g\beta}\delta_{\mu 4}+\frac{\partial_j\alpha(\vec{x})}{g}\delta_{\mu j}\,.
\end{equation}   
Thus in any admissible gauge the Polyakov loop $P$ on
$a_\mu^{D,\tiny\mbox{gs}}$ is unity:\\ 
$P[a_\mu^{D,\tiny\mbox{gs}}]=\exp\left[ig\int_0^\beta
  d\tau\,a_4^{D,\tiny\mbox{gs}}\right]=1$. 

The electric ${\mathbbm Z}_2$ degeneracy of the ground state, which
occured in the deconfining phase, no longer exists in the preconfining 
phase. Since the magnetic coupling $g$ remains finite inside this 
phase this does, however, not imply complete confinement 
since the dual gauge field, albeit massive, still propagates.    

In a way completely analogous to SU(2) one derives for 
SU(3) the following effective action for the preconfining phase
\cite{Hofmann2005L}:
\begin{equation}
\label{Spreconfsu3}
S=\sum_{l=1}^2\int_0^\beta d\tau\int d^3x\,\left[\frac14
  G^D_{\mu\nu,l}G^D_{\mu\nu,l}+\frac12 \left({\cal D}_{\mu,l}\varphi_l\right)^*{\cal
    D}_{\mu,l}\varphi_l+\frac12\frac{\bar{\Lambda}^6}{|\varphi_l|^2}\right]\,.
\end{equation}
Since SU(3)$\to$U(1)$^2$ in the deconfining 
phase there are now two independent species of magnetic 
monopoles, the dual gauge fields, $a_{\mu,1}^D$, $a_{\mu,2}^D$, and the monopole-antimonopole 
condensate, represented by inert complex scalar fields 
$\varphi_1$, $\varphi_2$. The magnetic coupling $g$ and the scale
$\bar{\Lambda}$ are universal, 
$a_{\mu,1}^{D,\tiny\mbox{gs}}=a_{\mu,2}^{D,\tiny\mbox{gs}}=\mp\frac{2\pi}{g\beta}\delta_{\mu 4}$, and the ground-state energy
density and pressure are given as
$\rho^{\tiny\mbox{gs}}=-P^{\tiny\mbox{gs}}=2\pi\bar{\Lambda}^3T$.
The Polyakov loop, evaluated on the ground-state configurations
$a_{\mu,1}^D$, $a_{\mu,2}^D$, 
is unity in any admissible gauge 
also for SU(3). This shows that the 
electric ${\mathbbm Z}_3$ degeneracy of the 
ground state of the deconfining phase no longer persists in the
preconfining phase.

\subsection{Thermal quasiparticle excitations of the dual gauge field}

\begin{proposition}
\label{QPEMu1}
In the effective theory for the preconfining phase the dynamical 
breaking of the residual gauge symmetry U(1) (for SU(2)) and U(1)$^2$ 
(for SU(3)) is manifested in terms of a quasiparticle mass $m$ for the dual
gauge field. One has $m=g|\varphi|=g|\varphi_1|=g|\varphi_2|=aT$ where
$a=2\pi g\bar{\lambda}^{-3/2}$.   
\end{proposition}
\begin{proof} 
In unitary gauge, $\varphi=|\varphi|=\varphi_{1,2}$ and 
$a_{\mu}^{D,\tiny\mbox{gs}}=a_{\mu,1}^{D,\tiny\mbox{gs}}=a_{\mu,2}^{D,\tiny\mbox{gs}}=0$,
the relation $m=g|\varphi|=g|\varphi_{1,2}|$ for the mass of
the fluctuations $\delta a_{\mu}^{D}$, $\delta a_{\mu,1}^{D}$, and 
$\delta a_{\mu,2}^{D}$ can be read off from (\ref{Spreconf}) and
(\ref{Spreconfsu3}), respectively (abelian Higgs mechanism), and $a=2\pi g\bar{\lambda}^{-3/2}$ then follows from 
Eq.\,(\ref{varphimod}) and the definition $\bar{\lambda}\equiv\frac{2\pi
T}{\bar{\Lambda}}$.  
\end{proof} 
\begin{remark}
Thus, the excitations in the effective theory for the deconfining phase
are {\sl free} thermal quasiparticles.  
\end{remark}
{\sl The contribution of quantum fluctuations to the thermodynamic 
pressure $\Delta V$ in the preconfining phase is
negligible.}\vspace{0.2cm}\\ 
For both SU(2) and SU(3) one obtains in close analogy to the deconfining
phase the following estimate
for the ratio $\frac{\Delta V}{V}$:
\begin{equation}
\label{qutotreeu1}
\left|\frac{\Delta V}{V}\right|\le\frac{\bar{\lambda}^{-3}}{24\pi^2}\,.
\end{equation}
As we shall see, $\bar{\lambda}\ge 7.075$ (SU(2)) and $\bar{\lambda}\ge
6.467$ (SU(3)). Thus $\Delta V$ is a small correction to the
(dominant\footnote{The total pressure $P$ is already negative at
  $\lambda_c$, see \cite{Hofmann2005L}.}) tree-level result $V$. 
\begin{proposition}
\label{scalematching}
The scales $\bar{\Lambda}$ (preconfining phase) and $\Lambda$
(deconfining phase) are related as
\begin{equation}
\label{smsu2su3}
\bar{\Lambda}=\left(4+\frac{\lambda_c^3}{720\pi^2}\right)^{1/3}\Lambda\,,\
\ \ (\mbox{for SU(2)})\,;\ \ \ 
\bar{\Lambda}=\left(2+\frac{\lambda_c^3}{720\pi^2}\right)^{1/3}\Lambda\,,\
\ \ (\mbox{for SU(3)})\,.
\end{equation}
\end{proposition}
\begin{proof} 
At $\lambda_c$, where $e=\infty$ and $g=0$, 
the pressure $P$ is continuous. Morever, 
no higher loop corrections to the one-loop result exist in 
the deconfining phase since in unitary gauge the fluctuations 
$a_{\mu}^{1,2}$ (SU(2)) and $a_{\mu}^{1,2,4,5,6,7}$ (SU(3)) decouple at $\lambda_c$. Equating at
$\lambda_c=\frac{\bar{\Lambda}}{\Lambda}\bar{\lambda}_c$ the right-hand side of
Eq.\,(\ref{Poneloop}) with the right-hand side of 
\begin{equation} 
\label{Prespreconf}
P(\bar{\lambda}_c)=-\bar{\Lambda}^4\left[\frac{6\bar{\lambda}_c^4}{(2\pi)^6}\bar{P}(0)+\frac{\bar{\lambda}_c}{2}\right]\,, \end{equation}
for the preconfining phase (negligible quantum
part), yields the claim for SU(2). 
For SU(3) one needs to equate the right-hand sides of 
\begin{equation} 
\label{Presdeeconfsu2out}
P(\lambda_c)=-\Lambda^4\left\{\frac{8\lambda_c^4}{(2\pi)^6}\bar{P}(0)+2\lambda_c\right\}\,
\end{equation}
and 
\begin{equation} 
\label{Prespreeconfsu2in}
P(\bar{\lambda}_c)=-\bar{\Lambda}^4\left\{\frac{12\bar{\lambda}_c^4}{(2\pi)^6}\bar{P}(0)+\bar{\lambda}_c\right\}\,.
\end{equation}
\end{proof} 
\begin{theorem}
\label{evolgsu2su3}
The evolution of the magnetic coupling $g$ with temperature is described
by the first-order differential equation
\begin{equation}
\label{evgsu2su3}
\partial_a\bar{\lambda}=-\frac{12\bar{\lambda}^4}{(2\pi)^6}\frac{a\,D(a)}{1+\frac{12\bar{\lambda}^3
    a^2}{(2\pi)^6}\,D(a)}\,,
\end{equation}
where $a\equiv 2\pi g\bar{\lambda}^{-3/2}$, and the
function $D(y)$ is defined below Eq.\,(\ref{evoleqsu2}). 
\end{theorem}
\begin{proof}
Because quantum contribution to the
pressure $P$ can be neglected in the effective theory for the
preconfining phase one has for SU(2
\begin{equation} 
\label{Prespreconfal}
P(\bar{\lambda})=-\bar{\Lambda}^4\left[\frac{6\bar{\lambda}^4}{(2\pi)^6}\bar{P}(a)+\frac{\bar{\lambda}}{2}\right]\,, 
\end{equation}
where the function $\bar{P}(y)$ is defined below
Eq.\,(\ref{rhooneloop}). The SU(3) pressure is just twice the SU(2)
pressure. As in the deconfining phase, the invariance of the Legendre transformations between
thermodynamic quantities under the applied coarse-graining implies for
the effective theory that $\partial_{(aT)} P=0$, and for both SU(2) and
SU(3) the same evolution equation (\ref{evgsu2su3}) follows.  
\end{proof}
\begin{remark}
\label{numprec}
Numerically, the initial condition for the evolution described by
Eq.\,(\ref{evgsu2su3}) is $g(\bar{\lambda}_c)=0$ for 
$\bar{\lambda}_c=8.478$ (SU(2)) and $\bar{\lambda}_c=7.376$ (SU(3)). 
For decreasing $\bar{\lambda}<\bar{\lambda}_c$ the magnetic coupling $g$
rises rapidly and runs into a logarithmic pole at $\bar{\lambda}_{c^\prime}$: $g\propto
-\log(\bar{\lambda}-\bar{\lambda}_{c^\prime})$. Numerically, one has
$\bar{\lambda}_{c^\prime}=7.075$ (SU(2)) and
$\bar{\lambda}_{c^\prime}=6.467$. Taking the mass $m$ of the dual 
gauge mode as an order parameter for the dynamical breaking of 
U(1) (SU(2)) and U(1)$^{2}$ (SU(3)) and postulating that 
$m=K(T_c-T)^\nu$ for $T\stackrel{<}\sim T_c$, where $K$ and $\nu$ are
constants, one extracts mean-field critical exponents: $\nu=\frac12$.   
\end{remark}
\begin{remark}
\label{energydensityprec}
The energy density $\rho$ divided by $T^4$ in the preconfining phase is given as
\begin{equation} 
\label{Prespreconfene}
\frac{\rho}{T^4}(\bar{\lambda})=\frac{(2\pi)^4}{\bar{\lambda}^4}
\left[\frac{6\bar{\lambda}^4}{(2\pi)^6}\bar{\rho}(a)+\frac{\bar{\lambda}}{2}\right]\,,\
  \ \ (\mbox{for SU(2)})\,;\ \ 
\frac{\rho}{T^4}(\bar{\lambda})=\frac{(2\pi)^4}{\bar{\lambda}^4}\left[\frac{12\bar{\lambda}^4}{(2\pi)^6}\bar{\rho}(a)+
\bar{\lambda}\right]\,,\
  \ \ (\mbox{for SU(3)})\,.
\end{equation}
\end{remark}
\begin{remark}
With a slight abuse of notation we refer to $\rho(\bar{\lambda})$ as the
functional dependence of the energy density on temperature
$\bar{\lambda}$ in the {\sl preconfining} phase and to $\rho(\lambda)$ as the
functional dependence of the energy density on temperature
$\lambda$ in the {\sl deconfining} phase. Thus $\rho(\bar{\lambda})$ and
$\rho(\lambda)$ are different functions of their arguments. 
\end{remark}
\begin{proposition}
\label{gapinen}
At $\lambda_c=\frac{\bar{\Lambda}}{\Lambda}\bar{\lambda}_c$ the energy
density $\rho$ exhibits a positive jump when decreasing the
temperature. One has
$\Delta(\bar{\lambda}_c)\equiv\frac{\rho(\lambda_c+0)-\rho(\bar{\lambda}_c-0)}{T_c^4}=\frac{4}{3}\,\frac{\pi^2}{30}$
for SU(2) and $\Delta(\bar{\lambda}_c)=\frac{8}{3}\,\frac{\pi^2}{30}$ for SU(3).
\end{proposition}
\begin{proof}
Routine computation considering Eqs.\,(\ref{smsu2su3}).
\end{proof}
\begin{remark}
\label{notreallyyet}
The existence of the gap $\Delta$ signals that the monopole-antimonopole condensate only builds
up gradually as temperature falls below $\lambda_c$. 
This is intuitively understandable because the 
condensation would require the influx of an infinite number of totally screened 
monopole-antimonopole pairs from infinity which costs energy. To 
facilitate the condensation of additional monopole-antimonopole pairs (stable
condensate) by total screening needs an increase of the average caloron/anticaloron holonomy from
almost trivial to maximal. Then the pairs of liberated monopoles and
antimonopoles screen one another, and no transport from infinity is
needed. Although this process is hard to grasp microscopically, 
after spatial coarse-graining the critical temperature at which a stable
condensate forms (defined by the property that the 
system is more likely to be preconfining than deconfining) can be determined exactly \cite{GH2005}. 
\end{remark}
\begin{remark}
\label{dofprec}
Notice that the number of degrees of freedom before coarse-graining
matches those after coarse graining. Namely, for SU(2) one has one species of
propagating gauge field times two polarizations plus 
one species of center-vortex loop, see Sec.\,\ref{CP}, before coarse-graining and one 
species of massive, dual gauge field times three polarization. Thus, 
one obtains three degrees of freedom {\sl before} and three degrees of
freedom {\sl after} coarse-graining. For SU(3) one obtains six 
degrees of freedom before and after coarse-graining.  
\end{remark}

\subsection{Supercooling\label{SC}} 

\begin{figure}[tb]
\centering
\includegraphics[width=11cm]{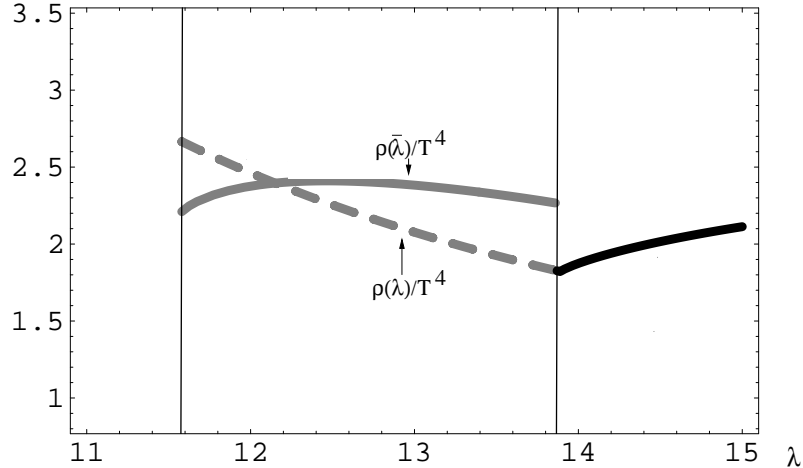}
\caption{\label{Fig-4B} The quantities $\frac{\rho(\lambda)}{T^4}$ (dashed line) and
  $\frac{\rho(\bar{\lambda})}{T^4}$ (solid grey line) in the preconfining
phase as functions of $\lambda$. The black solid line is associated with
$\frac{\rho(\lambda)}{T^4}$ in the deconfining phase.}
\end{figure}

{\sl A stable condensate of monopoles and antimonopoles exists for
temperatures $\bar{\lambda}$ with
$\bar{\lambda}_{c^\prime}\le\bar{\lambda}\le\bar{\lambda}_*$ where 
$\bar{\lambda}_{c^\prime}<\bar{\lambda}_*<\bar{\lambda}_{c}$.}\vspace{0.2cm}\\ 
At $\bar{\lambda}_{c^\prime}$, where $g=\infty$, one has 
$\frac{\rho(\bar{\lambda}_{c^\prime})}{T_{c^\prime}^4}=8\pi^4\,\bar{\lambda}^{-3}$
(for SU(2)) and
$\frac{\rho(\bar{\lambda}_{c^\prime})}{T_{c^\prime}^4}=16\pi^4\,\bar{\lambda}^{-3}$
(for SU(3)). On the other hand, continuing the energy density of the
deconfining phase down to
$\lambda_{c^\prime}=\frac{\bar{\Lambda}}{\Lambda}\bar{\lambda}_{c^\prime}$
and using Eqs.\,(\ref{smsu2su3}) yields
\begin{equation}
\label{condneedstohappen}
\frac{\rho(\lambda_{c^\prime})}{T_{c^\prime}^4}=\frac{\pi^2}{15}+\frac{32\pi^4}
{4+\frac{\lambda_c}{720\pi^2}}\,\bar{\lambda}_{c^\prime}^{-3}\,,\ \ \
(\mbox{for SU(2)})\,;\ \ \ 
\frac{\rho(\lambda_{c^\prime})}{T_{c^\prime}^4}=2\,\frac{\pi^2}{15}+\frac{32\pi^4}
{2+\frac{\lambda_c}{720\pi^2}}\,\bar{\lambda}_{c^\prime}^{-3}\,,\ \ \
(\mbox{for SU(3)})\,.
\end{equation} 
The second summands in Eqs.\,(\ref{condneedstohappen}) practically
coincide with the above expressions for 
$\frac{\rho(\bar{\lambda}_{c^\prime})}{T_{c^\prime}^4}$. 
Thus we conclude that 
$\frac{\rho(\bar{\lambda}_{c^\prime})}{T_{c^\prime}^4}<\frac{\rho(\lambda_{c^\prime})}{T_{c^\prime}^4}$
for both SU(2) and SU(3). But according to Prop.\,(\ref{gapinen}) 
we have
$\frac{\rho(\bar{\lambda}_{c})}{T_{c^\prime}^4}>\frac{\rho(\lambda_{c})}{T_{c}^4}$
for both SU(2) and SU(3). Since both functions
$\frac{\rho(\bar{\lambda})}{T^4}$ and 
$\frac{\rho(\lambda)}{T^4}$ are continuous in the ranges
$\bar{\lambda}_{c^\prime}\le\bar{\lambda}\le\bar{\lambda}_c$ and
$\lambda_{c^\prime}\le\lambda\le\lambda_c$, respectively, there is at
least one intersection. Numerically, one shows that only a single
intersection takes place, see Fig.\,\ref{Fig-4B}.
For SU(2) one obtains the following values: $\lambda_*=12.15$
or $\bar{\lambda}_*=7.428$. Now, cooling the system, 
a stable condensate (system is more likely to be found in preconfining than
in deconfining state) starts to take place at $\bar{\lambda}_*$, and the claim
follows.   
\begin{remark}
\label{stacoarsegraining}
The typical, maximal core-size $R_{\tiny\mbox{cor}}(\bar{\lambda})$ of an instable center-vortex 
loop is given as $R_{\tiny\mbox{cor}}(\bar{\lambda})\sim
\frac{1}{m}=\frac{1}{g|\varphi|}$ 
\cite{Hofmann2005L}. For
$\bar{\lambda}_{c^\prime}\le\bar{\lambda}\le\bar{\lambda}_*=7.428$,
where a stable monopole-antimonopole condensate exists, we have $g\ge 8.3$ according 
to Eq.\,(\ref{evgsu2su3}). Thus
$\frac{R_{\tiny\mbox{cor}}(\bar{\lambda})}{|\varphi|^{-1}}\le
0.12$. That is, collapsing center-vortex loops are not 
resolved\footnote{The resolution is given by $|\varphi|$.}, 
and the monopole-antimonopole condensate appears to be
spatially homogeneous.   
\end{remark}
\begin{remark}
\label{droplet}
To describe the average effect of tunneling between the two trajectories 
$\frac{\rho(\bar{\lambda})}{T^4}$ and $\frac{\rho(\lambda)}{T^4}$ for 
$\bar{\lambda}_*\le\bar{\lambda}\le\bar{\lambda}_c$ and
$\lambda_*\le\lambda\le\lambda_c$, respectively, one may think
of the following `droplet' model. Let $V\subset V_{\tiny\mbox{tot}}$ be two 
volumina. The thermal probability density $P(V,V_{\tiny\mbox{tot}},\bar{\lambda})$ for
measuring a fraction $\frac{V}{V_{\tiny\mbox{tot}}}$ of condensed magnetic monopoles 
and antimonopoles is given as 
\begin{equation}
\label{Porba} 
P(V,V_{\tiny\mbox{tot}},\bar{\lambda})\equiv
d(\bar{\lambda})\bar{\lambda}^3\,\frac{\exp\left[d(\bar{\lambda})\bar{\lambda}^3(V_{\tiny\mbox{tot}}-V)\right]}
{\exp\left[d(\bar{\lambda})\bar{\lambda}^3\,V_{\tiny\mbox{tot}}\right]-1}\,,
\end{equation}
where $d(\bar{\lambda})\equiv
\Delta(\bar{\lambda})\frac{\bar{\Lambda}^3}{(2\pi)^3}$, and $\Delta$ is
the temperature-dependent (positive) difference between 
trajectories $\frac{\rho(\bar{\lambda})}{T^4}$ and 
$\frac{\rho(\lambda)}{T^4}$. Notice that for
$d(\bar{\lambda})\bar{\lambda}^3 V_{\tiny\mbox{tot}}t\gg 1$ the probability density 
$P(V,V_{\tiny\mbox{tot}},\bar{\lambda})$ ceases to depend on $V_{\tiny\mbox{tot}}$. 
In the model of Eq.\,(\ref{Porba}) and for SU(2) 
the average polarization number $N_p$ of the U(1) gauge field calculates as
\begin{equation}
\label{N_p}
N_p(\bar{\lambda})=\int_0^{V_{\tiny\mbox{tot}}}
dV\,P(V,V_{\tiny\mbox{tot}},\bar{\lambda})\left(3\,\frac{V}{V_{\tiny\mbox{tot}}}+2\,\frac{V_{\tiny\mbox{tot}}-V}{V_{\tiny\mbox{tot}}}\right)=
2+\int_0^{V_{\tiny\mbox{tot}}}
dV\,P(V,V_{\tiny\mbox{tot}},\bar{\lambda})\,\frac{V}{V_{\tiny\mbox{tot}}}\,. 
\end{equation}
Keeping $V_{\tiny\mbox{tot}}$ fixed, this yields $\lim_{\bar{\lambda}\to\bar{\lambda}_*}
N_p=\lim_{d\to 0} N_p=\frac52$. A similar model for the regime
$\bar{\lambda}_{c^\prime}\le\bar{\lambda}\le\bar{\lambda}_*$ shows that 
$N_p$ increases towards $N_p=3$ for $\bar{\lambda}\searrow\bar{\lambda}_{c^\prime}$.  
\end{remark}

\section{Confining phase\label{CP}} 

In this section we turn to the confining phase which starts to set in at
the temperature $T_{c^\prime}$ where formerly instable, untwisted center-vortex 
loops become stable and massless \cite{Hofmann2005L} and the dual gauge
field decouples because its mass diverges, see
Rem.\,\ref{numprec}. At $T_{c^\prime}$ the magnetic ${\mathbbm Z}_2$
(for SU(2)) and ${\mathbbm Z}_3$ (for SU(3)) symmetries start to be
broken dynamically. Twisted or untwisted center-vortex loops, which are
liberated during the subsequent decay of the monopole-antimonopole condensate, are
interpreted as spin-1/2 fermions. 

The transition from
the preconfining to the confining phase is genuinely nonthermal: Relying
on the results of \cite{BenderWu1976} for the number of connected bubble
diagrams in a $\lambda\phi^4$ theory one proves by 
Borel summation and analytic continuation that the
pressure increasingly develops sign-indefinite imaginary parts as
temperature is increased from zero towards $T_{c^\prime}$ \cite{Hofmann2007}. One also
proves that at $T=0$ the pressure is precisely zero, see below \cite{Hofmann2007}. 

In the preconfining phase closed lines of magnetic flux\footnote{Since
  there are no isolated magnetic charges in the preconfining phase
  (monopoles and antimonopoles are condensed) these
flux lines cannot end and thus are closed.} form by the
collective dissociation of large-holonomy calorons/anticalorons. Inside 
their cores magnetic monopoles travel oppositely directed to 
their antimonopoles along the direction of the flux. The magnetic 
flux $F_{\pm,0}$ through a minimal spatial surface $A_C$, encircled by a close contour
$C$, is given as \cite{Hofmann2005L} 
\begin{equation}
\label{fluxC}
F_{\pm,0}=\left\{\begin{array}{c}\pm\frac{2\pi}{g}\\ 0\end{array}\right. 
\end{equation}
depending on whether a center-vortex flux 
pierces $A$ once ($\pm\frac{2\pi}{g}$) or whether it pierces $A_C$ not at
all or twice (0). Notice that $F_{\pm,0}$ does not depend on the velocity of 
the train of monopoles and antimonopoles travelling along the 
vortex line, for a discussion see \cite{Hofmann2005L}.
\vspace{0.2cm}\\ 
{\sl The phase of the dual order-parameter for confinement, the 't Hooft-loop
expectation (a complex field $\Phi$), 
takes on discrete values. These are $0,i\pi$ for SU(2) and
$0,\pm\frac{2\pi}{3}i$ for SU(3).}\vspace{0.2cm}\\    
The 't Hooft-loop operator $\hat{\Phi}(\vec{x},C)$ \cite{tHooft1978} 
is defined as the exponential of the magnetic flux 
of the dual gauge field $a^D_\mu$ through the minimal surface $M_C$ spanned by 
an oriented and closed spatial curve $C$ centered at the point $\vec{x}$:
\begin{equation}
\label{deftHooftloop}
\hat{\Phi}(\vec{x},C)\propto \exp[ig\oint_C dz_i\, a^D_i]\,.
\end{equation}
Thus the expectation $\Phi$ of $\hat{\Phi}$ changes phase 
under a singular gauge transformations of $a^D_\mu$ along the contour 
$C$ mediated by a local magnetic center jump. Such a center jump is associated 
with an extra quantum of center flux, induced by a 
center-vortex loop piercing $M_C$ in
addition the multitude of vortices which had generated the finite value 
of $\Phi$ to begin with. The process of having an additional vortex
pierce $M_C$ proceeds in real time, and it is clear that the smooth {\sl dynamics} of having 
$\Phi$ change its phase is described by a complex scalar 
field even though for SU(2) it is sufficient for the 
equilibrium situation to assume $\Phi$ to be a $Z_2$-charged real 
quantity. (Imagine the adiabatic limit, where no kinetic energy is associated
with the piercing center-vortex loop, no change in energy is conveyed to
$\Phi$ by this process. If $\Phi$ were real and the process were to be 
described by smooth dynamics then $\Phi$ would have to change its 
sign in a smooth way (exhibiting a zero in the process) 
which is in contradiction to energy conservation.) 

\begin{figure}[tb]
\centering
\includegraphics[width=11cm]{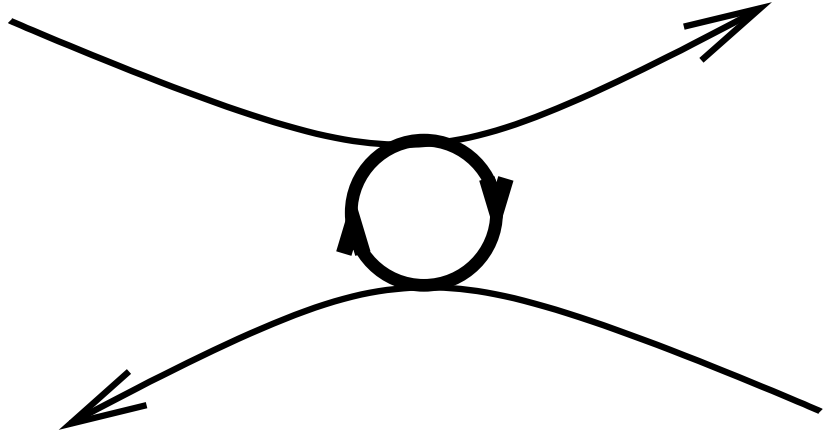}
\caption{\label{Fig-5}The core of a selfintersection in a center vortex loop.}
\end{figure}

\begin{figure}[tb]
\centering
\includegraphics[width=14cm]{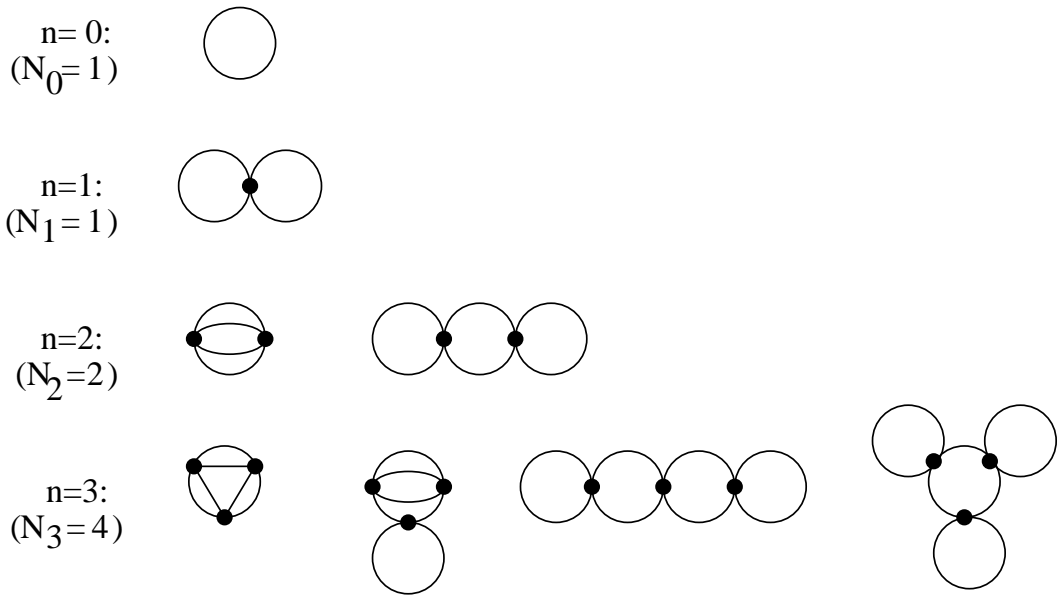}
\caption{\label{Fig-4}Untwisted and twisted center-vortex loops up to $n=3$}
\end{figure}

Consider now a spatial circle of infinite radius $S^{R=\infty}_1$ centered at
$\vec{x}$. The thermally averaged flux $F_{\pm,0;\tiny\mbox{th}}$ 
of a system of a center-vortex loop and its 
flux-reversed partner at rest through $A_{S^{R=\infty}_1}$ is in the limit of
vanishing core-size and mass
($\bar{\lambda}\to\bar{\lambda}_{c^\prime}$, $g\to \infty$) given 
as
\begin{equation}
\label{thermalCVL}
\lim_{g\to\infty}F_{\pm,0;\tiny\mbox{th}}=4\pi\int
d^3p\,\delta^{(3)}(\vec{p})\,n_B(\beta_{c^\prime} 2\,E_v(\vec{p},\bar{\lambda}_{c^\prime}))\,
F_{\pm,0}=\left\{\begin{array}{c}0\\ \pm\frac{\bar{\lambda}^{3/2}_{c^\prime}}{\pi}\,,\end{array}\right.
\end{equation}
where $E_v(0,\bar{\lambda})\sim \pi\frac{|\varphi(\bar{\lambda})|}{g}$ is the
typical mass of a single center-vortex loop at temperature 
$\bar{\lambda}$ \cite{Hofmann2005L}. Notice the use of the Bose function
$n_B$ for the system of two 
center-vortex loops of opposite-flux: Even though each vortex 
loop is interpreted as a spin-1/2 fermion (two polarizations also in the case
of selfintersections \cite{Hofmann2005L, Reinhardt2001}) the system 
is of spin zero. The value zero in
Eq.\,(\ref{thermalCVL}) is realized if $A_{S^{R=8}_1}$ is pierced an even
number of times by the center-vortex loops in the system, 
and the values
$\pm\frac{\bar{\lambda}^{3/2}_{c^\prime}}{\pi}$ 
correspond to odd numbers of piercings. It is obvious 
that in the SU(2) case an identification of 
$\pm\frac{\bar{\lambda}^{3/2}_{c^\prime}}{\pi}$ takes place which is not
true for SU(3). Properly normalized, the discrete values in
Eq.\,(\ref{thermalCVL}) are phase changes in $\Phi$ for the
creation of a single center-vortex loop. In SU(2) they are from 0 to $i\pi$ and from
$i\pi$ to 0. For SU(3) the process 0 to $\pm i\frac{2\pi}{3}$ and $\pm
i\frac{2\pi}{3}$ to 0 create two distinct species of center-vortex
loops. Since the spatial extent of a given center-vortex 
loop is unresolvable ($g\nearrow\infty$, for discussion see
\cite{Hofmann2005L}) and since in the condensate the distance
between a center-vortex loop and its flux-reversed partner is zero, 
the vanishing-curvature situation is trivially
saturated at finite curvature, $S^{R<\infty}_1$.\vspace{0.2cm}\\ 
{\sl The process of decay of the monopole-antimonopole condensate(s) and the 
formation of the center-vortex condensate is described by real-time 
dynamics of the order-parameter $\Phi$ subject 
to the potentials 
\begin{equation}
\label{potsCI}
V(\Phi)=\left(\frac{\tilde{\Lambda}^3}{\Phi}-\tilde{\Lambda}\Phi\right)^*
\left(\frac{\tilde{\Lambda}^3}{\Phi}-\tilde{\Lambda}\Phi\right)\,,\ \ \ \ 
(\mbox{for SU(2)})\,,
\end{equation}
\begin{equation}
\label{potsCII}
V(\Phi)=\left(\frac{\tilde{\Lambda}^3}{\Phi}-\Phi^2\right)^*
\left(\frac{\tilde{\Lambda}^3}{\Phi}-\Phi^2\right)\,,\ \ \ \ 
(\mbox{for SU(3)})\,,
\end{equation}
where $\tilde{\Lambda}\sim 2^{1/3}\,\bar{\Lambda}$ (for SU(2)) and
$\tilde{\Lambda}\sim\bar{\Lambda}$ (for SU(3)).}\vspace{0.2cm}\\ 
At the onset of center-vortex loop condensation thermal 
equilibrium is maintained at overall negative pressure. 
Periodic BPS saturated trajectories\footnote{A canonic 
kinetic term in $\Phi$'s effective action
\begin{equation}
\label{effacPhi}
S=\int d^4x\,\left(\frac12
  \left(\partial_\mu\Phi\right)^*\partial^\mu\Phi-\frac12 V\right)
\end{equation}
is inherited 
from the effective action for the field $\varphi$: At 
the onset of vortex condensation thermal equilibrium 
prevails, and the vortex condensate {\sl coincides} with the
monopole-antimonopole condensate.} along euclidean 
time, describing the onset of vortex-loop 
condensation, exist for the potentials in Eqs.\,(\ref{potsCI}) and
(\ref{potsCII}), see \cite{Hofmann2000}. As in the other two phases, this periodicity is 
due to the pole-term in the `square-root' of $V$ which endows the field
$\Phi$ with a winding number. (The `superpotential' ${\cal W}$ 
has a branch cut, see \cite{Hofmann2000}.) Furthermore, the potential
$V$ needs to satisfy the following requirements: (i) invariance under
(local) magnetic center transformation only (no larger symmetry), 
(ii) dynamical realization of the latter
(flux creation, negative tangential curvature for jump-like behavior in
real time), (iii) the only minima of $V$ are center-degenerate and
at zero energy density (center-vortex loops in condensate do not 
interact and are massless), and (iv) as in the other two phases, 
a single mass scale $\tilde{\Lambda}$ 
enters $V$. It is easy to check, see also \cite{Hofmann2005C}, that modulo U(1) invariant 
rescalings and adding term of the form $\Delta
V=\kappa\left(\tilde{\Lambda}^2-
\tilde{\Lambda}^{-2(n-1)}(\bar{\Phi}\Phi)^n\right)^{2k}\,,\ \ (\kappa>0,
k=1,2,3,\cdots, n\in{\mathbbm Z}$), which do increase the curvature 
of $V$ at its minima, the potentials in Eqs.\,(\ref{potsCI}) and
(\ref{potsCII}) are unique. Demanding at the onset of the condensation
of center-vortex loops that the (negative) pressure be 
continuous in the euclidean formulation, 
the above relation between the scales $\tilde{\Lambda}$ and
$\bar{\Lambda}$ follows.    
\begin{remark}
\label{grconf}
Writing $\Phi=|\Phi|\,\exp[i\frac{\theta}{\tilde{\Lambda}}]$, 
one has 
\begin{equation}
\label{minimacur}
\left.\frac{\pd^2_{\theta} V(\Phi)}{|\Phi|^2}\right|_{\Phi_{\tiny\mbox{min}}}=
\left.\frac{\pd^2_{|\Phi|} V(\Phi)}
{|\Phi|^2}\right|_{\Phi_{\tiny\mbox{min}}}
=\left\{\begin{array}{c}8\,\ \ \ \ \ (\mbox{SU(2)})\\ 
18\,\ \ \ \ \ (\mbox{SU(3)})\,,\end{array}\right.
\end{equation}
where $\Phi_{\tiny\mbox{min}}=\pm\tilde{\Lambda}$ (for SU(2)) and 
$\Phi_{\tiny\mbox{min}}=\tilde{\Lambda}\,\exp[\frac{2\pi i k}{3}]\,,\ \
(k=0,1,2)\,,$ (for SU(3)). Since $|\Phi|_{\tiny\mbox{min}}$ is the scale of maximal 
resolution once $\Phi$ has settled into one of its minima 
we conclude that the field $\Phi$ does no longer fluctuate. 
This, in turn, implies that no tunneling to another minimum
(flux creation) takes place once $\Phi$ has settled into
$\Phi_{\tiny\mbox{min}}$ \cite{Hofmann2005L}.      
\end{remark}
{\sl Naively, that is, without taking into account contact 
interactions between and internal excitations within (twisted)
center-vortex loops, the thermodynamic SU(2) pressure is estimated by the
following asymptotic-series representation\footnote{Apologies for
  introducing the variable $\lambda$ twice in this paper, here with a different meaning than in
  Secs.\,\ref{DP} and \ref{PP}.}
\label{asymptSer}
\begin{equation}
\label{estimaten>>1}
P_{\tiny\mbox{as}}\le 
\frac{M^4}{2\pi^2}{\hat{\beta}}^{-4}\left(\frac{7\pi^4}{180}+\sqrt{2\pi}\,{\hat{\beta}}^{\frac32}
\sum^L_{l=0}a_l\sum_{n\ge1}(32\lambda)^n\,n!\,n^{\frac{3}{2}+l}\right)\,,  
\end{equation}
where $\hat{\beta}\equiv\frac{M}{T}$, $M\sim\tilde{\Lambda}$, $\lambda\equiv \e^{-\hat{\beta}}$,
$L<\infty$, and $a_l\in {\mathbbm Z}$.}
\vspace{0.2cm}\\  
According to Rem.\,\ref{grconf} no (naive) contribution to 
the pressure arises from the ground state: The fermionic gas has
thermalized to a given temperature by the decay of the monopole-antimonopole 
condensate, and the field $\Phi$ is settled into one of the minima of
the potential $V$. Because of its two polarization states (two directions of center flux
for both SU(2) and SU(3) inherited from an untwisted progenitor
center-vortex loop) any center-vortex loop with $n$
selfintersections ($n=0,1,2,\cdots$) is interpreted as a spin-1/2
fermion. Its mass is $nM$ where $M$ corresponds to the mass of a single
intersection point (a ${\mathbbm Z}_2$ or a ${\mathbbm Z}_3$ monopole or 
antimonopole is associated with the core of the flux eddy, see
Fig.\,\ref{Fig-5}, 
marking the intersection -- a plastic
visualization of the concept of spin) which, in turn, is comparable to the scale
$\tilde{\Lambda}$. 
Since there are two possible charges of the monopole
singled out in the core of the intersection there are $C_n=2^n$ many
possible charge states of a center-vortex loop with $n$
selfintersections. Topologically, the multiplicity $N_n$ of these solitons is
known exactly up to $n=6$ \cite{KleinertSchulte-Frohlinde}, see
Fig.\,(\ref{Fig-4}).
For $n\gg 1$ the form $N_n\sim \left(\sum^L_{l=0}a_l\,n^l\right)n!\,16^n$ was obtained 
in \cite{BenderWu1976} by an analysis of the ground-state energy of the
anharmonic quantum mechanical oscillator ($\lambda\phi^4$-theory in one dimension). 
Naturally, $a_l\in{\mathbbm Z}$. Taking into account the spin
degeneracy, 
the total multiplicity $M_n$ of a center-vortex loop with $n$
selfintersections is given as 
\begin{equation}
\label{multCVLn}
M_n=2\times N_n\times C_n\stackrel{n\gg 1}=2\times 2^n\times\left(\sum^L_{l=0}a_l\,n^l\right)n!\,16^n\,.  
\end{equation}
Separating off the massless sector (single center-vortex loops) and
negelecting any interaction and internal excitability, one has 
\begin{equation}
P_{\tiny\mbox{as}}=\frac{M^4}{2\pi^2}{\hat{\beta}}^{-4}\left(\frac{7\pi^4}{180}+{\hat{\beta}}^{3}\sum_{n\ge1}M_n\int_0^\infty
dx\,x^2\,\log\left(1+\e^{-\hat{\beta}\sqrt{n^2+x^2}}\right)\right)\nonumber\\ 
\end{equation}
\begin{equation}
\le\frac{M^4}{2\pi^2}{\hat{\beta}}^{-4}\left(\frac{7\pi^4}{180}+{\hat{\beta}}^{3}\sum_{n\ge1}M_n\int_0^\infty
dx\,x^2\,\e^{-\hat{\beta}\sqrt{n^2+x^2}}\right)\nonumber\\ 
\end{equation}
\begin{equation}
=\frac{\Lambda^4}{2\pi^2}{\hat{\beta}}^{-4}\left(\frac{7\pi^4}{180}+{\hat{\beta}}^{2}
\sum_{n\ge1}M_n\,n^2\,K_2(n\hat{\beta})\right)\nonumber\\ 
\end{equation}
\begin{equation}
\sim\frac{M^4}{2\pi^2}{\hat{\beta}}^{-4}\left(\frac{7\pi^4}{180}+\sqrt{\frac{\pi}{2}}\,{\hat{\beta}}^{\frac32}
\sum_{n\ge1}M_n\,\lambda^n\,n^{\frac{3}{2}}\right)\nonumber\\ 
\end{equation}
\begin{equation}
\label{multCVLn5}
\le\frac{M^4}{2\pi^2}{\hat{\beta}}^{-4}\left(\frac{7\pi^4}{180}+\sqrt{2\pi}\,{\hat{\beta}}^{\frac32}
\sum^L_{l=0}a_l\sum_{n\ge1}(32\lambda)^n\,n!\,n^{\frac{3}{2}+l}\right)\,.
\end{equation}
Here $K_2$ denotes a modified Bessel function. In Eq.\,(\ref{multCVLn5}) the first $\le$ sign holds strictly for 
the linear truncation of the expansion of the logarithm about unity, and the $\sim$ sign 
indicates that terms of order $(\hat{\beta}n)^{-1}$ have been neglected
in the asymptotic expression for the Bessel function. 
This is relevant for studying the analyticity structure of the 
Borel resummed series. The second $\le$ sign holds because 
we have made use of the large-$n$ expression for $N_n$ of
Eq.\,(\ref{multCVLn}). 
Obviously, the expression in Eq.\,(\ref{estimaten>>1}) represents an
asymptotic series in $\lambda$ (zero radius of convergence). Upon sending 
$\lambda\to -\lambda$ in Eq.\,(\ref{multCVLn5}) notice the
formal similarity to the expansion of the ground-state energy of an 
anharmonic quantum mechanical oscillator \cite{BenderWu-1,BenderWu-2}
($\lambda\varphi^4$-theory in one dimension) for which Borel summability
was proven, see \cite{KleinertSchulte-Frohlinde} and refs. therein. 

The fact that the (naive) partition function diverges because of an
over-exponentially in energy rising density of states is known to be
associated with a so-called Hagedorn transition
\cite{Hagedorn1969}.\vspace{0.2cm}\\  
{\sl The asymptotic estimate in Eq.\,(\ref{estimaten>>1}) is Borel summable
for $\lambda<0$.}\vspace{0.2cm}\\ 
Notice that the case $\lambda<0$ corresponds to an analytic 
continuation from positive-real values of $\hat{\beta}\equiv
\hat{\beta}_1+i\hat{\beta}_2$ 
to complex values: $\hat{\beta}_2=0 \to \hat{\beta}_2=\pm\pi$.   

Let us now show the above claimed Borel summability. Defining $\bar{P}_{\tiny\mbox{mass}}(\bar{\lambda})\equiv
P_{\tiny\mbox{as}}-\frac{7\pi^2}{360}\left(\frac{M}{\hat{\beta}}\right)^{4}$
and\footnote{Apologies for introducing the variable $\bar{\lambda}$
  twice in this paper, here with a different meaning than in
  Sec.\,\ref{PP}.} $\bar{\lambda}\equiv 32\lambda$, the 
Borel transformation of $\bar{P}_{\tiny\mbox{mass}}(\bar{\lambda})$ is
given as 
\begin{equation}
\label{PtoB}
\bar{P}_{\tiny\mbox{mass}}(\bar{\lambda})\equiv\sum^L_{l=0}a_l\sum_{n\ge1}{\bar{\lambda}}^n\,n!\,n^{\frac{3}{2}+l}\ 
\ \ \overset{{\tiny \mbox{Borel}}}{\longrightarrow}
\ \ \ B_{\bar{P}_{\tiny\mbox{mass}}}(\bar{\lambda})\equiv\sum^L_{l=0}a_l\sum_{n\ge1} {\bar{\lambda}}^n\,n^{\frac{3}{2}+l}\,.
\end{equation}
Thus, $B_{\bar{P}_{\tiny\mbox{mass}}}(\bar{\lambda})$ is a superposition
of polylogarithms: 
\begin{equation}
\label{superpol}
B_{\bar{P}_{\tiny\mbox{mass}}}(\bar{\lambda})=\sum^L_{l=0}a_l\,  
\mbox{Li}_{-\left(\frac{3}{2}+l\right)}(\bar{\lambda})\,.
\end{equation}
The functions $\mbox{Li}_{-\left(\frac{3}{2}+l\right)}(\bar{\lambda})$
are real-analytic for $\bar{\lambda}<1$. To perform the inverse Borel 
transformation 
\begin{equation}
\label{inBoreltrafo}
\hat{P}_{\tiny\mbox{mass}}(\bar{\lambda})\equiv\sum_{l=0}^L a_l \hat{P}_l(\bar{\lambda})
\equiv\int_0^\infty dt\,\e^{-t}\,B_{\bar{P}_{\tiny\mbox{mass}}}(\bar{\lambda}\,t)\,,
\end{equation}
where 
\begin{equation}
\label{defpl}
\hat{P}_l(\bar{\lambda})\equiv\int_0^\infty dt\,\e^{-t}\,\mbox{Li}_{-\left(\frac{3}{2}+l\right)}(\bar{\lambda}\,t)\,,
\end{equation}
we notice the following
integral representation of $\mbox{Li}_{s}(z)$, valid for all
$s,z\in\mathbbm C$ \cite{Lewin}: 
\begin{equation}
\label{intrepLi}
\mbox{Li}_s(z)=\frac{iz}{2}\int_C du\,\frac{(-z)^u}{(1+u)^s\,\sin(\pi u)}\,,
\end{equation}
where the path $C$ is along the imaginary axis from $-i\infty$ to $+i\infty$ with 
an indentation to the left of the origin. Inserting Eq.\,(\ref{intrepLi}) into Eq.\,(\ref{defpl}) 
for $\bar{\lambda}=-|\bar{\lambda}|<0$ and interchanging the order of integration, 
we have 
\begin{equation}
\label{inverseBorel}
\hat{P}_l(\bar{\lambda})=-i\int_C du\,
\frac{(1+u)^{\frac32+l}}{1-\e^{-2\pi i\,u}}\,\e^{-\pi i\,u}\,\e^{(1+u)\log(-\bar{\lambda})}\,\Gamma(u+2)\,.
\end{equation}
Since, by Stirling's 
formula\footnote{$\Gamma(z)=\sqrt{2\pi}\, z^{z-\frac12}\,\e^{-z}\,e^{H(z)}$ where $H(z)\equiv\sum_{n\ge
0}\left(\left(z+n+1/2\right)\log\left(1+\frac{1}{z+n}\right)-1\right)$ 
converges for $z\in {\bf C}_{-}$ and $\lim_{|z|\to\infty} H(z)=0$, see \cite{Freitag}.} 
the gamma function $\Gamma(u+2)$ decays exponentially fast for $u\to\pm i\,\infty$ , the integral over $u$ 
in Eq.\,(\ref{inverseBorel}) exists and defines the real-analytic \footnote{This follows from Eq.\,(\ref{defpl}) and 
the fact that $\mbox{Im}\,\left[\mbox{Li}_{-\left(\frac{3}{2}+l\right)}(z)\right]\equiv 0$ for $z\le 0$.} 
function $\hat{P}_l(\bar{\lambda})\,,$ \ ($\bar{\lambda}<0$).  

The function $\hat{P}_l(\bar{\lambda})$ is analytic for a much larger 
range $\bar{\lambda}\in{\mathbbm C}$. Notice, however, that for
$\arg(\bar{\lambda})\to\pm\pi$ a branch cut is expected for 
$\hat{P}_l(\bar{\lambda})$. Also, one shows that $\hat{P}_l(0)=0$ by using 
$\e^{(1+u)\log(-\bar{\lambda})}=-\bar{\lambda}\e^{u\log(-\bar{\lambda})}$
in Eq.\,(\ref{inverseBorel}). 

The approximate behavior $\Phi_l(\bar{\lambda})$ of
$\hat{P}_l(\bar{\lambda})$ is suggested\footnote{One has:
\begin{equation}
\label{infit}
\int_0^{\infty}dx\,\e^{-ax}\,\sin(\log(-\bar{\lambda})x)=\frac{a}{a^2+(\log(-\bar{\lambda}))^2}\,,\
\
\int_0^{\infty}dx\,\e^{-ax}\,\cos(\log(-\bar{\lambda})x)=
\frac{\log(-\bar{\lambda})}{a^2+(\log(-\bar{\lambda}))^2}\,,\
\ \ (a>0)\,.\nonumber 
\end{equation}
} as follows:
\begin{equation}
\label{fitfunct}
\Phi_l(\bar{\lambda})=\frac{\sum_{r=0}^{R_l} \alpha_{2r+1}(\log(-\gamma_{2r+1}\bar{\lambda}))^{2r+1}}{\sum_{s=0}^{S_l} 
\beta_{2s}(\log(-\delta_{2s}\bar{\lambda}))^{2s}}\,,
\end{equation}
where $\gamma_{2r+1},\delta_{2s}\in {\mathbbm R}_+ $, 
$\alpha_{2r+1},\beta_{2s}\in {\mathbbm R}$ and $S_l=R_l+1$. 
Numerically, one has for example
\begin{equation}
\Phi_0(\bar{\lambda})=0.0570\frac{\log(-0.154\bar{\lambda})}{1+0.220(\log(-0.494\bar{\lambda}))^2}\,,\nonumber\\ 
\end{equation}
\begin{equation}
\label{fitfunctexe2}
\Phi_1(\bar{\lambda})=\frac{0.0212\log(-10.2\bar{\lambda})+0.00142(\log(-0.109\bar{\lambda}))^3}
{1+0.128(\log(-1.09\bar{\lambda}))^2+0.0544(\log(-0.886\bar{\lambda}))^4}\,.
\end{equation}
Since for $\Phi_l$ one has $\Phi_l(\bar{\lambda}=0)=0$ due to a higher 
power of the logarithmic singularity in the numerator than in 
the denominator it is clear that $|\mbox{Im}\Phi_l|$ grows slower than 
$|\mbox{Re}\Phi_l|$ for sufficiently small, real-positive values of $\bar{\lambda}$ 
increasing from zero. Also, $\mbox{Re}\Phi_l$ is continuous across the branch
cut. The growing importance of imaginary 
contaminations of the physical pressure with increasing 
temperature signals the growing deviation from genuine thermal 
equilibrium: A sign-indefinite imaginary part implies the existence 
of exponentially fast growing and decaying plasma modes and thus
turbulences.    

Since the only difference for SU(3) is the occurrence of two types of
center-vortex loops one obtains the result for the latter by simply
multiplying the SU(2) result by two.  

\section{Conclusion}

A detailed discussion and partial analysis of the thermodynamics of SU(2) and SU(3)
Yang-Mills theory has been given. As for the case of SU(2) there appears to be a wealth of 
applications in particle physics \cite{Hofmann2005L,Hofmann2005S}, 
cosmology \cite{Hofmann2005-1,GH2005,SHG2006-1,SHG2006-2}, 
and plasma physics \cite{Hofmann2007,ZPinch} (dark energy by virtue of
the axial anomaly, leptons and their interactions). To enable future contact with
experiment in the case of SU(3) (strong interactions) the dynamics of electric-magnetically 
dual gauge-group factors (fractional quantum Hall effect) 
needs to be explored in their confining phases.

\subsection*{Acknowledgments}
The author would like to acknowledge joyful 
collaboration with Francesco Giacosa on some aspects of the
here-presented material. I am also grateful for 
excellent work conducted by my students. Many thanks go to Nucu Stamatescu 
for interesting discussions, and to Markus Schwarz and Francesco Giacosa
for their helpful comments on the manuscript. I am indebted to my family, 
in particular to my wife Karin Thier, 
for their persistent help and understanding over the last four years. 
I also would like to thank Frans Klinkhamer for suggesting 
the present paper and for having the far-sight and fairness 
to care about my institutional survival. Finally, I would like to thank
the Referee for a very thorough assessment of the manuscript and 
various helpful suggestions for improvement.             

\LastPageEnding

\end{document}